# Sophisticated Inference

Karl Friston[1], Lancelot Da Costa[1,2], Danijar Hafner[3,4], Casper Hesp[1,5], Thomas Parr[1]

[1]*The Wellcome Centre for Human Neuroimaging, UCL Queen Square Institute of Neurology, London, UK*

[2]*Department of Mathematics, Imperial College London, UK*

[3]*Department of Computer Science, University of Toronto, Toronto, Canada*

[4]*Google Research, Brain Team, Toronto, Canada*

[5]*Amsterdam Brain and Cognition Center (ABC), University of Amsterdam, The Netherlands*

Emails: k.friston@ucl.ac.uk, l.da-costa@imperial.ac.uk, mail@danijar.com, c.hesp@uva.nl, thomas.parr.12@ucl.ac.uk

**Correspondence**: Thomas Parr

*The Wellcome Centre for Human Neuroimaging*

*Institute of Neurology, UCL*

*12 Queen Square, London, UK WC1N 3AR*

thomas.parr.12@ucl.ac.uk

## Abstract

Active inference offers a first principle account of sentient behaviour, from which special and important cases can be derived, e.g., reinforcement learning, active learning, Bayes optimal inference, Bayes optimal design, *etc*. Active inference resolves the exploitation-exploration dilemma in relation to prior preferences, by placing information gain on the same footing as reward or value. In brief, active inference replaces value functions with functionals of (Bayesian) beliefs, in the form of an expected (variational) free energy. In this paper, we consider a sophisticated kind of active inference, using a recursive form of expected free energy. Sophistication describes the degree to which an agent has beliefs about beliefs. We consider agents with beliefs about the counterfactual consequences of action for states of affairs *and* beliefs about those latent states. In other words, we move from simply considering beliefs about "what would happen if I did that" to "what would I *believe about* what would happen if I did that". The recursive form of the free energy functional effectively implements a deep tree search over actions and outcomes in the future. Crucially, this search is over sequences of belief states, as opposed to states *per se*. We illustrate the competence of this scheme, using numerical simulations of deep decision problems.

# Introduction

In theoretical neurobiology, active inference has proved useful in providing a generic account of motivated behaviour under ideal Bayesian assumptions, incorporating both epistemic and pragmatic value (Da Costa et al., 2020a; Friston et al., 2017a). This account is often portrayed as being based on first principles, because it inherits from the statistical physics of random dynamical systems at nonequilibrium steady-state (Friston, 2013; Hesp et al., 2019a; Parr et al., 2020). Active inference does not pretend to replace existing formulations of sentient behaviour – it just provides a Bayesian mechanics from which most (and, arguably, all) normative optimisation schemes can be derived as special cases. Generally, these special cases arise when ignoring one sort of uncertainty or another. For example, if we ignore uncertainty about (unobservable) hidden states that generate (observable) outcomes, active inference reduces to conventional schemes like optimal control theory and reinforcement learning. While the latter schemes tend to focus on the maximisation of value as a function of hidden states *per se*, active inference optimises a functional[1] of (Bayesian) beliefs *about* hidden states. This allows it to account for uncertainties surrounding action and perception in a unified, Bayes-optimal fashion.

Most current applications of active inference rest on the selection of policies (i.e., ordered sequences of actions; or open-loop policies, where the sequence of future actions depends only on current states but not future states) that minimise a functional of beliefs called *expected free energy* (Da Costa et al., 2020a; Friston et al., 2017a). This approach clearly has limitations, in the sense that one has to specify *a priori* allowable policies, each of which represents a possible path through a deep tree of action sequences. This formulation limits the scalability of the ensuing schemes, because only a relatively small number of policies can be evaluated (Tschantz et al., 2019). In this paper, we consider active inference schemes that enable a deep tree search over all allowable sequences of action into the future. Because this involves a recursive evaluation of expected free energy – and implicit Bayesian beliefs – the resulting scheme has a sophisticated aspect (Costa-Gomes et al., 2001; Devaine et al., 2014); namely, rolling out beliefs about beliefs.

---

[1] Technically, a 'functional' is defined as a function whose arguments (in this case, beliefs about hidden states) are themselves functions of other arguments (in this case, observed outcomes generated by hidden states).

'Sophistication' is a term from the economics literature and refers to having beliefs about one's own or another's beliefs. For instance, in game theory, an agent is said to have a level of sophistication of one if she has beliefs about her opponent; two if she has beliefs about her opponent's beliefs about her strategy; and so forth. Most people have a level of sophistication greater than two (Camerer et al., 2004).

On this view, most current illustrations of active inference can be regarded as unsophisticated or naive, in the sense that they only consider beliefs about the consequences of action, as opposed to the consequences of action for beliefs. In what follows, we try to unpack this distinction intuitively and formally, using mathematical and numerical analyses. We also take the opportunity to survey the repertoire of existing schemes that fall under the Bayesian mechanics of active inference; including, expected utility theory (Von Neumann and Morgenstern, 1944), Bayesian decision theory (Berger, 2011), optimal Bayesian design (Lindley, 1956), reinforcement learning (Sutton and Barto, 1981), active learning (MacKay, 1992), risk sensitive control (van den Broek et al., 2010), artificial curiosity (Schmidhuber, 2006), intrinsic motivation (Oudeyer and Kaplan, 2007), empowerment (Klyubin et al., 2005), and the information bottleneck method (Tishby et al., 1999; Tishby and Polani, 2010).

Sophisticated inference recovers Bayes-adaptive reinforcement learning (Åström, 1965; Ghavamzadeh et al., 2016; Ross et al., 2008) in the zero temperature limit. Both approaches perform belief state planning, where the agent maximizes an objective function by taking into account how it expects its own beliefs to change in the future (Duff, 2002), and evinces a degree of sophistication. The key distinction is that Bayes-adaptive reinforcement learning considers arbitrary reward functions, while sophisticated active inference optimises an expected free energy that can be motivated from first principles. While both can be specified for particular tasks, the expected free energy additionally mandates the agent to seek out information about the world (Friston, 2013; Friston, 2019) beyond what is necessary for solving a particular task (Tishby and Polani, 2010). This allows inference to account for artificial curiosity (Lindley, 1956; Oudeyer and Kaplan, 2007; Schmidhuber, 1991), that goes beyond reward seeking to the gathering of evidence for an agent's existence (i.e., its marginal likelihood). This is sometimes referred to as self-evidencing (Howhy, 2016).

The basic distinction between sophisticated and unsophisticated inference was briefly introduced in Appendix 6 of (Friston et al., 2017a). This distinction is subtle but can lead to fundamentally different kinds of behaviour. We will use a simple example to illustrate the difference: consider the following three-armed bandit problem – with a twist. The right and left arms increase or decrease your winnings. However, you do not know which arm is which. The central arm does not affect your winnings but tells you which arm pays off. Crucially, once you have committed to either the right or the left arm, you cannot switch to the other arm. This game is engineered to confound agents who choice behaviour is based upon Bayesian decision theory. This follows because the expected payoff is the same for every sequence of moves. In other words, choosing the right or left arm – for the first and subsequent trials – means you are equally likely to win or lose. Similarly, choosing the middle arm (or indeed doing nothing) has the same Bayesian risk or expected utility.

However, an active inference agent – who is trying to minimise her expected free energy[2] – will select actions that minimise the risk of losing and resolve her uncertainty about whether the right or left arm pays off. This means that the centre arm acquires epistemic (uncertainty-resolving) affordance and becomes intrinsically attractive. On choosing the central arm – and discovering which arm holds the reward – her subsequent choices are informed, in the sense that she can exploit her knowledge and commit to the rewarding arm. In this example, the agent has resolved a simple exploration-exploitation dilemma[3], by resolving ambiguity as a prelude to exploiting updated beliefs about the consequences of subsequent action. Note that having selected the central arm there is no ambiguity in play and its epistemic affordance disappears. Note further that, initially, all three arms have some epistemic affordance; however, the right and left arms are less informative if the payoff is probabilistic.

---

[2] Expected free energy can be read as risk plus ambiguity: *risk* is taken here to be the relative entropy (i.e., KL divergence) between predicted and preferred outcomes, while *ambiguity* is the conditional entropy (i.e., conditional uncertainty) about outcomes given their causes.

[3] Exploration here has been associated with the resolution of ambiguity or uncertainty about hidden states; namely, the context in which the agent is operating (i.e., left, or right arm payoff). More conventional formulations of exploration could remove the prior belief that the right and left arms have a complementary payoff structure, such that the agent has to learn the probabilities of winning and losing, when selecting either arm. However, exactly the same principles apply: the right and left arms now acquire an epistemic affordance in virtue of resolving uncertainty about the contingencies that underlie payoffs – as opposed to hidden states. We will see how this falls out of expected free energy minimisation later.

The key move behind this paper is to consider a sophisticated agent who evaluates the expected free energy of each move recursively. Simply choosing the central arm to resolve uncertainty does not, in and of itself, mean an epistemic action was chosen in the service of securing future rewards. In other words, the central arm is selected because all the options had the same Bayesian risk[4]; while the central arm had the greatest epistemic affordance[5]. Now consider a sophisticated agent, who imagines what she will do after acting. For each plausible outcome, she can work out how her beliefs about hidden states will be updated – and evaluate the expected free energy of the subsequent move, under each action and subsequent outcome. By taking the average over both, she can evaluate the expected free energy of the second move that is afforded by the first. If she repeats this process recursively, she can, effectively, perform a deep tree search over all ordered sequences of actions and their consequences.

Heuristically, the unsophisticated agent simply chooses the central arm because he knows it will resolve uncertainty about states of affairs. Conversely, the sophisticated agent follows through – on this resolution of ambiguity – in terms of its implications for subsequent choices. In this instance, she knows that only two things can happen if she chooses the central arm; either the right or left arm will be disclosed as the payoff arm. In either case, the subsequent choice can be made unambiguously to minimise risk and secure her reward. The average expected free energy of these subsequent actions will be pleasingly low, making a choice of the central arm more attractive than its expected free energy would otherwise suggest. This means the sophisticated agent is more confident about her choices because she has gone beyond forming beliefs about the consequences of action to consider the effects of action on subsequent beliefs and the (epistemic) actions that ensue. The remainder of this paper unpacks this recursive kind of planning, using formal analysis and simulations.

This paper comprises three sections. The first provides a brief overview of active inference in terms of free energy minimisation and the various schemes that can be used for implementation. This section starts with the basic imperative to optimise Bayesian beliefs about latent or hidden states of the world in terms of approximate Bayesian (i.e., variational) inference (Dayan et al., 1995). It then goes on to cast planning as inference (Attias, 2003; Botvinick and Toussaint, 2012) as the minimisation of an expected free energy under allowable sequences of actions or policies (Friston et al., 2017a). The foundations of expected free energy are detailed in an appendix from two complementary perspectives, the second of which is probably more fundamental as it rests upon the first-principle account mentioned above (Friston, 2013; Friston, 2019; Parr et al., 2020). The second section considers sophisticated schemes using a recursive formulation of expected free energy. Effectively, this enables the efficient search of deep policy trees (that entail all possible outcomes under each policy or path). This search is efficient because only those paths that have a sufficiently high predictive posterior probability need to be evaluated. This restricted tree search is straightforward to implement in the present setting because we are propagating beliefs (i.e., probabilities) – as opposed to value functions. The third section provides some illustrative simulations that compare sophisticated and unsophisticated agents in the three-armed bandit (or T-maze paradigm) described above. This final section considers deeper problems, using navigation and novelty seeking as an example. We conclude with a brief summary of what sophisticated inference brings to the table.

---

[4] Bayesian risk is taken to be negative expected utility, i.e., expected loss under some predictive posterior beliefs (about hidden states).

[5] Epistemic affordance is taken to be the information gain or relative entropy of predictive beliefs (about hidden states), before and after an action.

## Active inference and free energy minimisation

Most of the active inference literature concerns itself with partially observable Markov decision processes. In other words, it considers generative models of discrete hidden states and observable outcomes, with uncertainty about the (likelihood) mapping between hidden states and outcomes – and (prior) probability transitions among hidden states. Crucially, sequential policy selection is cast as an inference problem, by treating sequences of actions (i.e., policies) as random variables. Planning then simply entails optimising posterior beliefs about the policies being pursued – and selecting an action from the most likely policy.

On this view, there are just two sets of unknown variables: namely, hidden states and policies. Belief distributions over this bipartition can then be optimised with respect to an evidence bound in the usual way, using an appropriate mean-field approximation (Beal, 2003; Winn and Bishop, 2005). In this setup, we can associate *perception* with the optimisation of posterior beliefs about *hidden states*, while *action* follows from planning based upon posterior beliefs about *policies*. Implicit in this formulation is a generative model; namely, a probabilistic specification of the joint probability distribution over policies, hidden states, and outcomes. This generative model is usually factorised into the likelihood of outcomes, given hidden states, the conditional distribution over hidden states, given policies, and prior over policies. In active inference, the priors over policies is determined by their expected free energy, noting that the expected free energy, which depends upon future courses of action, furnishes an empirical prior over subsequent actions.

In brief, given some prior beliefs about the initial and final states of some epoch of active inference, the game is to find a posterior belief distribution over policies that brings the initial distribution as close as possible to the final distribution, given observations. This objective can be achieved by optimising posterior beliefs about hidden states and policies with respect to a variational bound on (the logarithm of) the marginal likelihood of the generative model (i.e., log evidence). This evidence bound is known as a variational free energy, or (negative) evidence lower bound. In what follows, we overview the formal aspects of this enactive kind of inference.

## Discrete state-space models

Our objective is to optimise beliefs (i.e., an approximate posterior) over policies $\pi$ and their consequences; namely, hidden states $s \equiv s_{\leq \tau}$ from some initial state $s_1$, until some policy horizon $\tau$, given some observations $o_{\leq t}$ up until the current time $t$. This optimisation can be cast as minimising a (generalised) free energy functional $F[Q(s, \pi)]$ of the approximate posterior (Parr and Friston, 2019b). This generalised free energy has two parts: the first entails a generative model for state transitions, given policies, while the second entails a generative model for policies that depend upon the final states (omitting constants for clarity):

$$F[Q(s, \pi)] = \mathbb{E}_{Q(\pi)}[F(\pi)] + D_{KL}[Q(\pi) \| P(\pi)]$$
$$= \mathbb{E}_{Q(\pi)}[\ln Q(\pi) + E(\pi) + F(\pi) + G(\pi)]$$

$$F(\pi) = \mathbb{E}_{Q(s_{\leq \tau} | \pi)}[\ln Q(s_{\leq \tau} | \pi) - \ln P(o_{\leq t}, s_{\leq \tau} | \pi)] \tag{1.1}$$
$$G(\pi) = \mathbb{E}_{Q(o_\tau, s_\tau | \pi)}[\ln Q(s_\tau | \pi) - \ln P(o_\tau, s_\tau)]$$

$$Q(o_\tau, s_\tau | \pi) = P(o_\tau | s_\tau) Q(s_\tau | \pi)$$
$$- \ln P(\pi) = E(\pi) + G(\pi)$$

This generalised free energy includes the *variational* free energy of each policy $F(\pi)$ that depends on priors over state transitions – and an *expected* free energy of each policy $G(\pi)$ that underwrites priors over policies. The priors over policies $\ln P(\pi) = -E(\pi) - G(\pi)$ ensure the expected free energy at time $\tau$ (i.e., the policy horizon) is minimised. Here, $E(\pi)$ represents an empirical prior that is usually conditioned upon hidden states at a higher level in deep (i.e., hierarchical) generative models. Note that outcomes on the horizon are random variables with a likelihood distribution, whereas outcomes in the past are realised variables.

The first equality above shows that the variational free energy, expected under the posterior over policies, plays the role of an accuracy, while the complexity of posterior beliefs about policies is the divergence from prior beliefs[6]. In other words, variational free energy scores the evidence for a particular policy that accrues from observed outcomes. The priors over policies also have the form of a free energy. For interested readers, the appendix provides a fairly comprehensive motivation of this functional form, from complementary perspectives. We now consider the role of free energy in exact, approximate, and amortized inference – before turning to active inference and policy selection.

## Perception as inference

Optimising the posterior over hidden states renders the variational free energy equivalent to (negative) log evidence – or marginal likelihood – in the usual way, while optimising the posterior over policies renders the generalised free energy zero:

$$Q\left(s \mid \pi\right) = \arg\min_{Q} F(\pi) = P(s_{\leq t} \mid o_{\leq t}, \pi)$$
$$\Rightarrow F(\pi) = -\ln P(o_{\leq t} \mid \pi)$$

$$Q(\pi) = \arg\min_{Q} F[Q(s, \pi)]$$
$$= \sigma[-E(\pi) - F(\pi) - G(\pi)]$$
$$\Rightarrow F[Q(s, \pi)] = 0$$

$$(1.2)$$

The first equalities correspond to *exact Bayesian inference* based on a softmax function – i.e., normalised exponential, $\sigma[\cdot]$ – of the log probability over outcomes and hidden states, under a particular policy. To finesse the numerics of optimising the posterior over all hidden states, a mean-field approximation usually leverages the Markovian form of the generative model to optimise an approximate posterior over hidden states at each time point (where $s_{\backslash \tau}$ denotes the Markov blanket of $s_\tau$):

$$Q\left(s_\tau \mid \pi\right) = \sigma[\mathbb{E}_{Q(s_{\backslash \tau} \mid \pi)}[\ln P(o_{\leq t}, s_{\leq \tau} \mid \pi)]]$$
$$= \sigma[\mathbb{E}_{Q(s_{\backslash \tau} \mid \pi)}[\ln P(o_\tau \mid s_\tau) + \ln P(s_\tau \mid s_{\tau-1}, \pi) + \ln P(s_{\tau+1} \mid s_\tau, \pi)]]$$

$$(1.3)$$

$$Q\left(s \mid \pi\right) = Q(s_1 \mid \pi)Q(s_2 \mid \pi)\ldots Q(s_\tau \mid \pi)$$
$$P\left(s \mid \pi\right) = P(s_1 \mid \pi)P(s_2 \mid s_1, \pi)\ldots P(s_\tau \mid s_{\tau-1}, \pi)$$

---

[6] Generally speaking, log evidence is accuracy minus complexity, where accuracy is the expected log likelihood and complexity is the KL divergence between posterior and prior beliefs.

This corresponds to a form of *approximate Bayesian inference* (i.e., variational Bayes) in which (1.3) is iterated over the factors of the mean-field approximation, to perform a coordinate descent or fixed-point iteration (Beal, 2003). An alternative formulation rests on an explicit minimisation of variational free energy using iterated gradient flows to each fixed point (expressed in terms of sufficient statistics):

$$\dot{\mathbf{v}}_\tau^\pi = -\partial_{\mathbf{s}_\tau^\pi} F(\pi) = \mathbb{E}_{Q(s_{\backslash \tau} \mid \pi)}[\ln P(o_{\leq t}, s_{\leq \tau} \mid \pi)] - \ln Q(s_\tau \mid \pi)$$
$$\mathbf{s}_\tau^\pi = \sigma(\mathbf{v}_\tau^\pi)$$

$$(1.4)$$

$$Q\left(s_\tau \mid \pi\right) = Cat(\mathbf{s}_\tau^\pi)$$

This solution can be read as (neuronal) dynamics that implement variational message passing[7] (Beal, 2003; Friston et al., 2017c; Parr et al., 2019). In this form, the free energy gradients constitute a prediction error; namely, the difference between the posterior surprisal[8] and its predicted value.

Finally, one can consider *amortizing inference* using standard procedures from machine learning to optimise the parameters $\phi$ of a recognition model w.r.t. variational free energy. In the present setting, this approach can be summarised as using universal function approximators (e.g., deep neural networks) to parameterise (1.2); namely, the mapping between observations and the sufficient statistics of the approximate posterior. For example,

$$\mathbf{s}_\tau^\pi = f_\phi(o_{\leq t}, \mathbf{s}_{\leq \tau}^\pi, \pi)$$
$$\phi = \arg\min_\phi F[Q(s, \pi)] \qquad (1.5)$$
$$Q\left(s_\tau \mid \pi\right) = Cat(f_\phi)$$

---

[7] Where $v$ can be thought of as transmembrane voltage or depolarisation and $s$ corresponds to the average firing rate of a neuronal population. Da Costa, L., Parr, T., Sengupta, B., Friston, K., 2020b. Natural selection finds natural gradient. arXiv arXiv:2001.08028.

[8] Surprisal is the self-information or negative log probability of outcomes. Tribus, M., 1961. Thermodynamics and Thermostatics: An Introduction to Energy, Information and States of Matter, with Engineering Applications. D. Van Nostrand Company Inc, New York, USA.

Effectively, amortized inference is 'learning to infer' (Catal et al., 2019; Lee and Keramati, 2017; Millidge, 2019; Toussaint and Storkey, 2006; Tschantz et al., 2019; Ueltzhöffer, 2018). Variational autoencoders can be regarded as an instance of amortized inference, if we ignore conditioning on policies: e.g., (Suh et al., 2016). Clearly, amortization precludes online inference and, as such, may appear biologically implausible. However, it might be the case that certain brain structures learn to infer: e.g., the cerebellum might learn from inferential processes implemented by the cerebral cortex (Doya, 1999; Ramnani, 2014).

## Planning as inference

The posterior over policies is somewhat simpler to evaluate – as a softmax function of their empirical [9], variational and expected free energy. This can be expressed in terms of a generalised free energy that includes the parameters of the generative model (e.g., the likelihood parameters, $A$):

$$Q(\pi) = \arg\min_Q F[Q(s, \pi, A)] = \sigma[-E(\pi) - F(\pi) - G(\pi)]$$
$$G(\pi) = \mathbb{E}_{Q(o_\tau, s_\tau | \pi)Q(A)}[\ln Q(s_\tau | \pi)Q(A) - \ln P(o_\tau, s_\tau, A)] \tag{1.6}$$

The expected free energy of a policy can be unpacked in a number of ways. Perhaps the most intuitive is in terms of risk and ambiguity[10]:

$$G(\pi) = \underbrace{D_{KL}[Q(s_\tau, A | \pi) \| P(s_\tau, A)]}_{\text{Risk}} + \underbrace{\mathbb{E}_{Q(o_\tau, s_\tau | \pi)}[-\ln P(o_\tau | s_\tau, A)]}_{\text{Ambiguity}} \tag{1.7}$$

This means that policy selection minimises *risk* and *ambiguity*. Risk, in this setting, is simply the difference between predicted and prior beliefs about final states. In other words, policies will be deemed more likely if they bring about states that conform to prior preferences. In the optimal control literature, this part of expected free energy underwrites KL control (Todorov, 2008; van den Broek et al., 2010). In economics, it leads to risk sensitive policies (Fleming and Sheu, 2002). Ambiguity reflects the uncertainty about future outcomes, given hidden states. Minimising ambiguity therefore corresponds to choosing future states that generate unambiguous and informative outcomes (e.g., switching on a light in the dark).

Sometimes, it is useful to express risk in terms of outcomes, as opposed to hidden states. For example, when the generative model is unknown – or one can only quantify preferences about outcomes (as opposed to the inferred causes of those outcomes). In these cases, the risk over hidden states can be replaced risk over outcomes, by assuming the divergence between the predicted and true posterior is small (omitting parameters for clarity):

---

[9] The empirical free energy is usually based upon inferences at a higher level in a hierarchical generative model. For details on hierarchical generative models, please see Friston, K.J., Rosch, R., Parr, T., Price, C., Bowman, H., 2017d. Deep temporal models and active inference. Neuroscience and biobehavioral reviews 77, 388-402.

[10] The appendix provides derivations of (1.7) based upon the principles of optimal Bayesian design and an integral fluctuation theorem described in Friston, K., 2019. A free energy principle for a particular physics, eprint arXiv:1906.10184..

$$\underbrace{D_{KL}[Q(s_\tau \,|\, \pi) \,\|\, P(s_\tau)]}_{\text{Risk (states)}} = \underbrace{D_{KL}[Q(o_\tau \,|\, \pi) \,\|\, P(o_\tau)]}_{\text{Risk (outcomes)}} + \underbrace{\mathbb{E}_{Q(o_\tau|\pi)}[D_{KL}[Q(s_\tau \,|\, o_\tau, \pi) \,\|\, P(s_\tau \,|\, o_\tau)]}_{\text{Expected evidence bound}} \qquad (1.8)$$

This divergence constitutes an expected evidence bound that also appears if we unpack expected free energy in terms of *intrinsic* and *extrinsic value*[11]:

$$\begin{aligned}
\mathrm{G}(\pi) = &-\underbrace{\mathbb{E}_{Q(o_\tau|\pi)}[\ln P(o_\tau)]}_{\text{Extrinsic value}} + \underbrace{\mathbb{E}_{Q(o_\tau|\pi)}[D_{KL}[Q(s_\tau, A \,|\, o_\tau, \pi) \,\|\, P(s_\tau, A \,|\, o_\tau)]]}_{\text{Expected evidence bound}} \\
&- \underbrace{\mathbb{E}_{Q(o_\tau|\pi)}[D_{KL}[Q(s_\tau \,|\, o_\tau, \pi) \,\|\, Q(s_\tau \,|\, \pi)]]}_{\text{Intrinsic value (states) or salience}} - \underbrace{\mathbb{E}_{Q(o_\tau, s_\tau|\pi)}[D_{KL}[Q(A \,|\, o_\tau, s_\tau, \pi) \,\|\, Q(A)]]}_{\text{Intrinsic value (parameters) or novelty}}
\end{aligned} \qquad (1.9)$$

$$\geq -\underbrace{\mathbb{E}_{Q(o_\tau|\pi)}[\ln P(o_\tau)]}_{\textit{Expected log evidence}} - \underbrace{\mathbb{E}_{Q(o_\tau|\pi)}[D_{KL}[Q(s_\tau, A \,|\, o_\tau, \pi) \,\|\, Q(s_\tau, A \,|\, \pi)]]}_{\textit{Expected information gain}}$$

Extrinsic value is just the expected value of log prior preferences (i.e., log evidence), which can be associated with reward and utility in behavioural psychology and economics, respectively (Barto et al., 2013; Kauder, 1953; Schmidhuber, 2010). In this setting, extrinsic value is the complement of Bayesian risk (Berger, 2011). The intrinsic value of a policy is its epistemic value or affordance (Friston et al., 2015). This is just the expected information gain afforded by a particular policy, which can be about hidden states (i.e., *salience*) or model parameters (i.e., *novelty*). It is this term that underwrites artificial curiosity (Schmidhuber, 2006). The final inequality above shows that extrinsic value is the expected log evidence under beliefs about final outcomes, while the intrinsic value ensures that this expectation is maximally informed, when outcomes are encountered. Collectively, these two terms underwrite the resolution of uncertainty about hidden states (i.e., information gain) and outcomes (i.e., expected surprisal) in relation to prior beliefs.

Intrinsic value is also known as intrinsic motivation in neurorobotics (Barto et al., 2013; Oudeyer and Kaplan, 2007; Ryan and Deci, 1985), the value of information in economics (Howard, 1966), salience in the visual neurosciences and (rather confusingly) Bayesian surprise in the visual search literature (Itti and Baldi, 2009; Schwartenbeck et al., 2013; Sun et al., 2011). In terms of information theory, intrinsic value is mathematically equivalent to the expected mutual information between hidden states in the future and their consequences – consistent with the principles of minimum redundancy or maximum efficiency (Barlow, 1961; Barlow, 1974; Linsker, 1990). Finally, from a statistical perspective, maximising intrinsic value (i.e., salience and novelty) corresponds to optimal Bayesian design (Lindley, 1956) and machine learning derivatives, such as active learning (MacKay, 1992). On this view, active learning is driven by novelty; namely, the information gain afforded to beliefs about model parameters, given future states and their outcomes. Heuristically, this curiosity resolves uncertainty about "what would happen if I did that?" (Schmidhuber, 2010). Figure 1 illustrates the compass of expected free energy, in terms of its special cases, ranging from optimal Bayesian design through to Bayesian decision theory.

---

[11] Because the expected evidence bound cannot be less than zero, the expected free energy of a policy is always greater than the (negative) expected intrinsic value (i.e., log evidence) plus the intrinsic value (i.e., information gain).

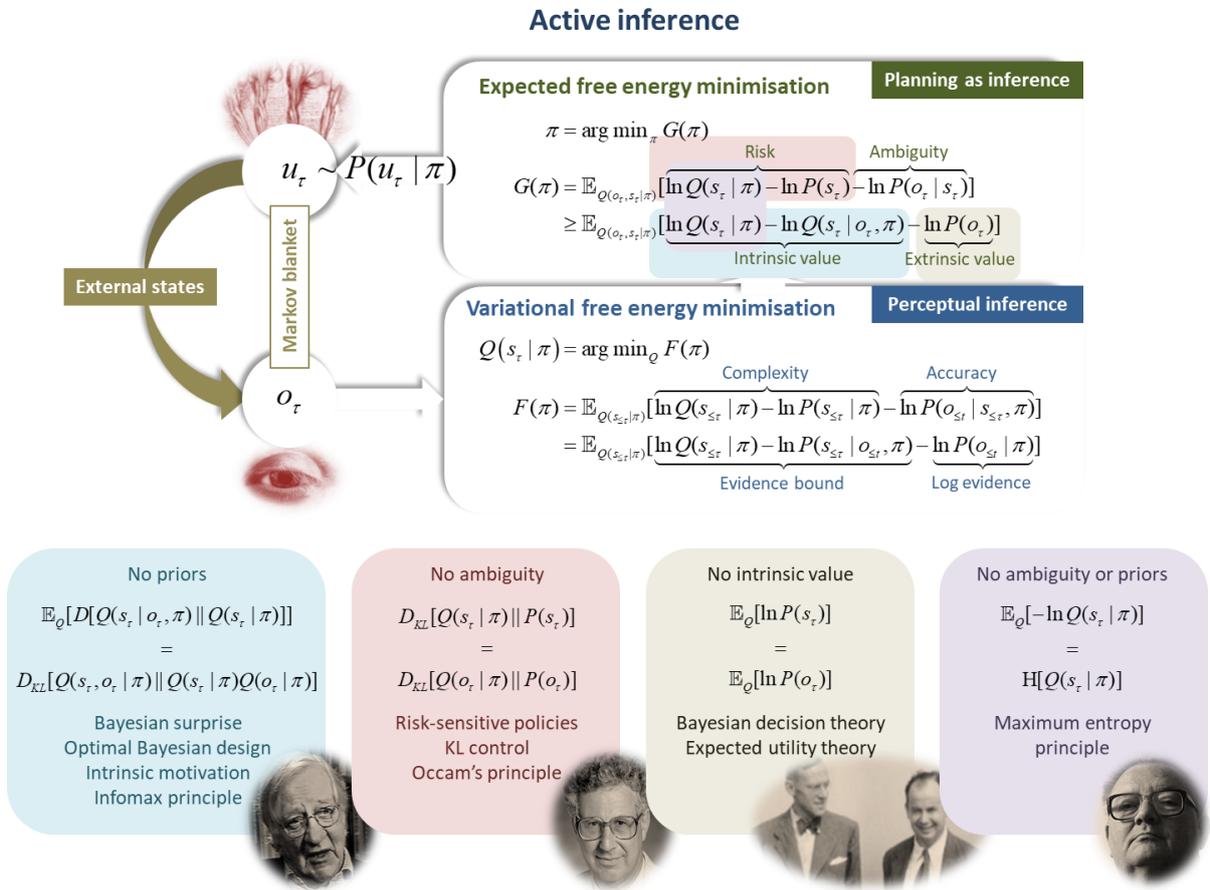

**FIGURE 1**

**Active inference:** This figure illustrates the various ways in which minimising expected free energy can be unpacked. The upper panel casts action and perception as the minimisation of variational and expected free energy, respectively. Crucially, active inference introduces beliefs over policies that enable a formal description of planning as inference (Attias, 2003; Botvinick and Toussaint, 2012; Kaplan and Friston, 2018). In brief, posterior beliefs about hidden states of the world, under plausible policies, are optimised by minimising a variational (free energy) bound on log evidence. These beliefs are then used to evaluate the expected free energy of allowable policies, from which actions can be selected (Friston et al., 2017a). Crucially, expected free energy subsumes several special cases that predominate in psychology, machine learning and economics. These special cases are disclosed when one removes particular sources of uncertainty from the implicit optimisation problem. For example, if we ignore prior preferences, then the expected free energy reduces to information gain (Lindley, 1956; MacKay, 2003) or intrinsic motivation (Barto et al., 2013; Oudeyer and Kaplan, 2007; Ryan and Deci, 1985). This is mathematically the same as expected Bayesian surprise and mutual information that underwrites salience in visual search (Itti and Baldi, 2009; Sun et al., 2011) and the organisation of our visual apparatus (Barlow, 1961; Barlow, 1974; Linsker, 1990; Optican and Richmond, 1987). If we now remove risk but reinstate prior preferences, one can effectively treat hidden and observed (sensory) states as isomorphic. This leads to risk sensitive policies in economics (Fleming and Sheu, 2002; Kahneman and Tversky, 1979) or KL control in engineering (van den Broek et al., 2010). Here, minimising risk corresponds to aligning predicted outcomes to preferred outcomes. If we then remove intrinsic value, we are left with extrinsic value or expected utility in economics (Von Neumann and Morgenstern, 1944) that underwrites reinforcement learning and behavioural psychology (Sutton and Barto, 1998). Bayesian formulations of maximising expected utility under uncertainty is also known as Bayesian decision theory (Berger, 2011). Finally, if we just consider a completely unambiguous world with uninformative priors, expected free energy reduces to the negative entropy of posterior beliefs about the causes of data; in accord with the maximum entropy principle (Jaynes, 1957). The expressions for variational and expected free energy correspond to those described in the main text (omitting model parameters for clarity). They are arranged to illustrate the relationship between *complexity* and *accuracy*, which become *risk* and *ambiguity*, when considering the consequences of action. This means that risk-sensitive policy selection minimises expected complexity or computational cost (Sengupta and Friston, 2018).

## Sophisticated inference

So far, we have considered generative models of policies; namely, a fixed number of ordered action sequences. These generative models can be regarded as placing priors over actions that stipulate a small number of allowable action sequences. In what follows, we consider more general models, in which the random variables are actions at each point in time; such that policies become a prior over transitions among action or control states. If we relax this prior, such that successive actions are conditionally independent, we can simplify belief updating – and implicit planning – at the expense of having to consider a potentially enormous number of policies.

The simplification afforded by assuming actions are conditionally independent follows because both actions and states become Markovian. This means we can use belief propagation (Winn and Bishop, 2005; Yedidia et al., 2005) to update posterior beliefs about hidden states and actions, given each new observation. In other words, we no longer need to evaluate the posterior over hidden states in the past to evaluate a posterior over policies. Technically, this is because policies introduced a semi-Markovian aspect to belief updating, by inducing conditional dependencies between past and future hidden states. The upshot of this is that one can use posterior beliefs from the previous time step as empirical priors for hidden states and actions at the subsequent time step. This is formally equivalent to the forward pass in the forwards-backwards algorithm (Ghahramani and Jordan, 1997), where the empirical prior over hidden states depends upon the preceding (i.e., realised) action. Put simply, we are implementing a Bayesian filtering scheme in which observations are generated by action at each time step. Crucially, the next action is sampled from an empirical prior based upon (a free energy functional of) posterior beliefs about the current hidden state.

Note that we do not need to evaluate a posterior over action, because action is realised before the next observation is generated. In other words, we can sample realised actions from an empirical prior over actions that inherits from the posterior over all previous states. This leads to a simple belief-propagation scheme for planning as inference that can be expressed as follows:

$$Q\left(s_\tau \mid u_{<\tau}\right) = P(s_\tau \mid o_{<\tau}, u_{<\tau}) = \mathbb{E}_{Q(s_{\tau-1})}[P(s_\tau \mid s_{\tau-1}, u_{\tau-1})]$$

$$Q(s_\tau) = P(s_\tau \mid o_{\leq\tau}, u_{<\tau}) \propto P(o_\tau \mid s_\tau) Q(s_\tau \mid u_{<\tau})$$

$$Q(u_\tau) = \sigma[G(u_\tau)]$$

(1.10)

$$G(u_\tau) = \mathbb{E}_{P(o_{\tau+1}\mid s_{\tau+1})Q(s_{\tau+1}\mid u_{<\tau+1})}[\underbrace{\ln Q(s_{\tau+1} \mid u_{<\tau+1}) - \ln P(s_{\tau+1})}_{\text{Risk}} \underbrace{- \ln P(o_{\tau+1} \mid s_{\tau+1})}_{\text{Ambiguity}}]$$

$$\underbrace{\qquad\qquad\qquad\qquad\qquad\qquad\qquad\qquad\qquad\qquad\qquad}_{\text{Expected free energy of next action}}$$

Here, $Q(s_\tau \mid u_{<\tau})$ denotes an empirical prior – from the point of view of state estimation – or a predictive posterior – from the point of view of action selection – over hidden states, given realised actions $u_{<\tau}$. Similarly, $Q(s_\tau)$ denotes the corresponding posterior, given subsequent outcomes. This scheme is exact because we have made no mean-field approximations of the sort required by variational message passing (Dauwels, 2007; Friston et al., 2017c; Parr et al., 2019; Winn and Bishop, 2005). Furthermore, there are no conditional dependencies on policies, which have been replaced by realised actions. However, equation (1.10) only considers the next action. The question now arises: how many future actions should we consider?

At this point the cost of the Markovian assumption arises: if we choose a policy horizon that is too far into the future, the number of policies could be enormous. In other words, we could effectively induce a deep tree search over all possible sequences of future actions that would be computationally prohibitive. However, we can now turn to sophisticated schemes to finesse the combinatorics. This rests upon the straightforward observation that if we propagate beliefs and uncertainty into the future, we only need to evaluate policies or paths that have a nontrivial likelihood of being pursued. This selective search over plausible paths is constrained at two levels. First, by propagating probability distributions, we can restrict the search over future outcomes – for any given action at any point in the future – that have a non-trivial posterior probability (e.g., greater than 1/16). Similarly, we only need to evaluate those policies that are likely to be pursued; namely, those with an expected free energy that renders their prior probability nontrivial (e.g., greater than 1/16).

This deep search involves evaluating all actions under all plausible outcomes, so that one can perform counterfactual belief updating at each point in time (given all plausible outcomes). However, it is not necessary to evaluate outcomes *per se* – it is sufficient to evaluate distributions over outcomes, conditioned upon plausible hidden states. This is a subtle but important aspect of finessing the combinatorics of belief propagation into the future – that rests upon having a generative model (that generates outcomes).

Heuristically, one can imagine searching a tree with diverging branches at successive times in the future but terminating the search down any given branch when the prior probability of an action (and the predictive posterior probability of its subsequent outcome) reach a suitably small threshold (Keramati et al., 2016; Solway and Botvinick, 2015). To form a marginal empirical prior over the next action, one simply accumulates the average expected free energy from all the children of a given node in the tree recursively. A softmax function of this accumulated average then constitutes the empirical prior over the next action. Algorithmically, this can be expressed as follows – based on Appendix 6 (Friston et al., 2017a) – where $u_\tau$ denotes action at $\tau \geq t$ (omitting novelty terms associated with model parameters for clarity):

$$G(o_\tau, u_\tau) = \mathbb{E}_{P(o_{\tau+1}|s_{\tau+1})Q(s_{\tau+1}|u_{<\tau+1})} [\underbrace{\ln Q(s_{\tau+1} \mid u_{<\tau+1}) - \ln P(s_{\tau+1})}_{\text{Risk}} \underbrace{- \ln P(o_{\tau+1} \mid s_{\tau+1})}_{\text{Ambiguity}}]$$

$$\underbrace{\phantom{G(o_\tau, u_\tau)}}_{\text{Expected free energy of next action}}$$

$$+ \underbrace{\mathbb{E}_{Q(u_{\tau+1}|o_{\tau+1})Q(o_{\tau+1}|u_{\leq\tau})} [G(o_{\tau+1}, u_{\tau+1})]}_{\text{Expected free energy of subsequent actions}}$$

$$(1.11)$$

$$Q(u_\tau \mid o_\tau) = \sigma[G(o_\tau, u_\tau)]$$
$$Q(o_\tau \mid u_{<\tau}) = \mathbb{E}_{Q(s_\tau|u_{<\tau})}[P(o_\tau \mid s_\tau)]$$

Posterior beliefs over hidden states and empirical priors over action are then recovered from the above recursion as follows, noting that one's most recent action $(u_{t-1})$ and current outcome $(o_t)$ are realised (i.e., known) variables:

$$Q(s_t) \propto P(o_t \mid s_t) Q(s_t \mid u_{<t})$$
$$Q(u_t) = \sigma[G(o_t, u_t)]$$
(1.12)

Equation (1.11) expresses the expected free energy of each potential next action $(u_t)$ as the risk and ambiguity of that action plus the average expected free energy of future beliefs, under counterfactual outcomes and actions $(u_{t+1})$. Readers familiar with the Bellman optimality principle (Bellman, 1952) may recognise a formal similarity between (1.11) and the Bellman equation because both inherit from the same recursive logic. The sophisticated inference scheme deals with functionals (functions of belief distributions over states), while the Bellman equation deals directly with functions of states.

Figure 2 provides a schematic that casts this recursive formulation as a deep tree search. This search can be terminated at any depth or horizon. Later, we will rewrite this recursive scheme in terms of sufficient statistics to illustrate its simplicity. Having established the formal basis of sophisticated planning, in terms of belief propagation, we now turn to some illustrative examples to show how it works in practice.

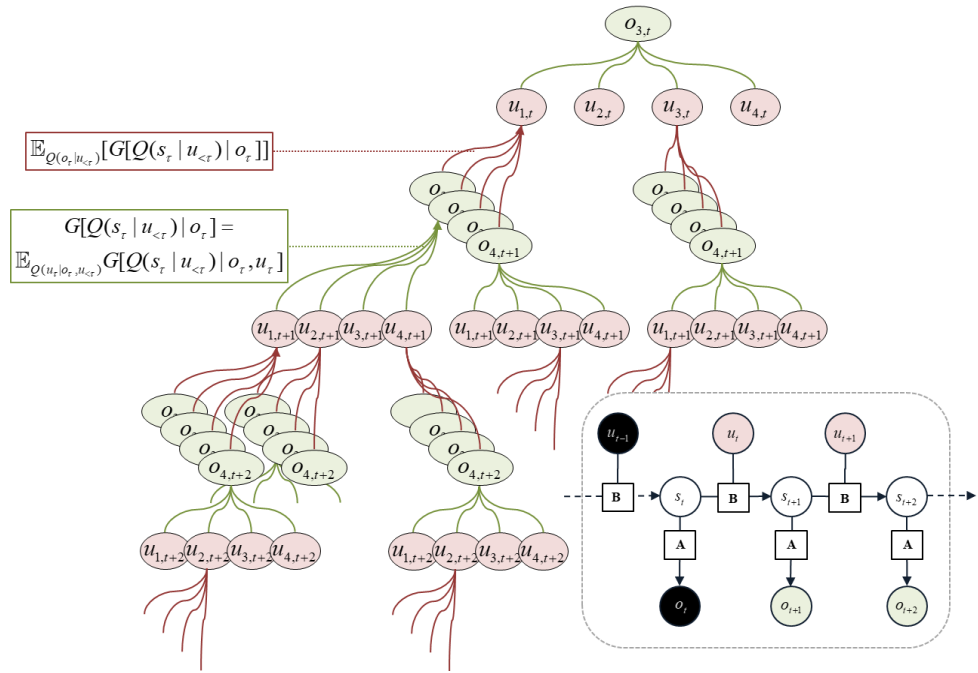

**FIGURE 2**

**Deep policies searches**. This schematic summarises the accumulation of expected free energy over paths or trajectories into the future. This can be construed as a deep tree search, where the tree branches over allowable actions at each point in time – and the likely outcomes consequent upon each action. The arrows between actions and outcomes have been drawn in the reverse direction (directed from the future) to depict the averaging of expected free energy over actions (green arrows) and subsequent averaging over the outcomes entailed by the preceding action (pink arrows). This dual averaging over actions (given outcomes) and outcomes (given actions) is depicted by the equations in the upper panel. Here, the green nodes of this tree correspond to outcomes, with one (realised) outcome at the current time (at the top). The pink nodes denote actions; here, just four. Note that the search terminates whenever an action is deemed unlikely – or an outcome is implausible. The panel on the lower right represents the conditional dependencies in the generative model as a probabilistic graphical model. The parameters of this model are shown on squares, whereas the variables are shown on circles. The arrows denote conditional dependencies. Filled circles are realised variables at the current time; namely, the preceding action and the subsequent outcome.

## Simulations

In this section, we provide some simulations to compare sophisticated and unsophisticated schemes on the three-arm bandit task described in the introduction. Here, we will frame this paradigm in terms of a rat, foraging in a three arm T-maze, where the right and left upper arms are baited with rewards and punishments, while the bottom arm contains an instructional cue indicating whether the bait is likely to be on the right or left. In these examples, cue validity was 95%. The details of this setup have been described elsewhere (Friston et al., 2016; Friston et al., 2017a). In brief, the generative model comprises a likelihood mapping between hidden states and outcomes – and probability transitions among states. Here, there are two outcome modalities: the first reports the experience of the rat in terms of its location (with distinct outcomes for the instructional cue location – *right* versus *left*). The second modality registered rewarding outcomes, with three levels (*none*, *reward,* and *punishment* – e.g., foot shock). There were two hidden factors: the rat's location (with four possibilities) and the latent context (i.e., whether the rewarding arm was on the *right* or the *left*). With these hidden states and outcomes, we specify the generative model in terms of:

- The sensory mapping **A**, which maps from the two hidden state factors (location and context) to each of the two sensory modalities (location and reward).
- The transition matrices **B**, which govern how states at one time point map onto the next, given a particular action ($u_t$). The transitions among locations are action-dependent, with four actions (moving to one of the four locations), while the context did not change during any particular trial (i.e., there were no context transitions within trials).
- The cost vectors **C** for each hidden state factor, which also specify the agent's preferences for each outcome modality. The latter allows for an alternative formulation that we discuss below.
- The priors over initial states, **D**.

In the following simulations, the rat experienced 32 trials, each comprising two moves with three outcomes, including an initial outcome that located the rat at the start (i.e., centre) location. The rat encountered the first trial with ambiguous prior beliefs about the context, i.e., the reward was equally likely to be right or left.

Given this parameterisation of the generative model, the expected free energy of an action, given outcomes, (1.11) can be expressed in terms of sufficient statistics of posterior beliefs and model parameters as follows[12]:

---

[12] We have suppressed any tensor notation here by assuming there is only one outcome modality and one hidden factor. In practice, this assumption can be guaranteed by working with the Kronecker tensor product of hidden factors. This ensures exact Bayesian inference, because conditional dependencies among hidden factors are evaluated.

$$\mathbf{G}(u_\tau, o_\tau) = \underbrace{\mathbf{s}_{\tau+1}^u \cdot [\ln \mathbf{s}_{\tau+1}^u + \mathbf{C} + \mathbf{H}]}_{\text{Expected free energy of next action}} + \underbrace{\mathbf{u}_{\tau+1}^o \cdot \mathbf{G}(u_{\tau+1}, o_{\tau+1}) \mathbf{o}_{\tau+1}^u}_{\text{and subsequent actions}} \tag{1.13}$$

$$\mathbf{s}_\tau \propto (\mathbf{A} \cdot o_\tau) \odot \mathbf{s}_\tau^u$$
$$\mathbf{s}_\tau^u = \mathbf{B}(u_{\tau-1}) \mathbf{s}_{\tau-1}$$
$$\mathbf{o}_\tau^u = \mathbf{A} \mathbf{s}_\tau^u$$
$$\mathbf{u}_\tau^o = \sigma[\mathbf{G}(u_\tau, o_\tau)]$$

Here, $\odot$ denotes a Hadamard (i.e., element-wise) product and the dot notation means $\mathbf{A} \cdot o_\tau \equiv \mathbf{A}^T o_\tau$. The sufficient statistics are the parameters of the categorical distributions in (1.11), where model parameters are usually hyper-parameterised in terms of the concentration parameters of Dirichlet distributions (denoted by capital and small bold variables, respectively):

$$Q(s_\tau) = Cat(\mathbf{s}_\tau)$$
$$Q(s_{\tau+1} \mid u_{\leq\tau}) = Cat(\mathbf{s}_{\tau+1}^u)$$
$$Q(u_\tau \mid o_\tau) = Cat(\mathbf{u}_\tau^o)$$
$$Q(o_\tau \mid u_{<\tau}) = Cat(\mathbf{o}_\tau^u)$$

$$P(o_\tau \mid s_\tau) = Cat(\mathbf{A})$$
$$P(s_{\tau+1} \mid s_\tau, u_\tau) = Cat(\mathbf{B}(u_\tau))$$
$$P(s_1) = Cat(\mathbf{D})$$
$$\mathbf{C} = -\ln P(s_\tau)$$
$$\mathbf{H} = -diag(\mathbf{A} \cdot \ln \mathbf{A})$$

$$P(\mathbf{A}) = Dir(a)$$
$$P(\mathbf{B}) = Dir(b)$$
$$P(\mathbf{D}) = Dir(d) \tag{1.14}$$

The equivalent scheme, when specifying preferences in terms of outcomes $\mathbf{C} = \ln P(o_\tau)$ is:

$$\mathbf{G}(u_\tau, o_\tau) = \underbrace{\mathbf{o}_{\tau+1}^u \cdot [\ln \mathbf{o}_{\tau+1}^u + \mathbf{C}] + \mathbf{s}_{\tau+1}^u \cdot \mathbf{H}}_{\text{Expected free energy of next action}} + \underbrace{\mathbf{u}_{\tau+1}^o \cdot \mathbf{G}(u_{\tau+1}, o_{\tau+1}) \mathbf{o}_{\tau+1}^u}_{\text{and subsequent actions}} \tag{1.15}$$

As noted above, it is usually more convenient to search over distributions over outcomes that are generated by (plausible) hidden states – as opposed to (plausible) outcomes *per se*. This approach produces a slightly simpler form for expected free energy:

$$\mathbf{G}(u_\tau, o_\tau) = \mathbf{s}_{\tau+1}^u \cdot [\underbrace{\ln \mathbf{s}_{\tau+1}^u + \mathbf{C} + \mathbf{H}}_{\text{Next action}} + \underbrace{\mathbf{G}(u_{\tau+1}, \mathbf{A}\mathbf{s}_{\tau+1}^u) \cdot \mathbf{u}_{\tau+1}^o}_{\text{and subsequent actions}}] \tag{1.16}$$

Finally, as intimated above, the recursive estimation of expected free energy from subsequent states can be terminated, when the probability of an action or outcome can be plausibly discounted. In the simulations here, searches over paths were terminated when the predictive probability fell below 1/16.

The simulations were chosen to illustrate the fidelity of beliefs about action (i.e., what to do next) with and without a sophisticated update scheme; i.e., (1.10) and (1.11). We anticipated that sophisticated schemes would outperform unsophisticated schemes, in the sense that they would learn any contingencies more efficiently, via more confident action selection. This learning was elicited by baiting the left arm consistently, after a couple of trials – so that priors about the initial (latent context) state could be accumulated, in the form of posterior (Dirichlet) concentration parameters ($\mathbf{d}$). In these generative models, learning is straightforward and involves the accumulation of posterior concentration parameters (Friston et al., 2016). For example, to learn the likelihood mapping and initial hidden states, we have[13]:

$$\mathbf{A} = \mathbf{a} \odot \mathbf{a}_0^{\odot -1}, \quad \mathbf{a}_{0ij} = \sum_i \mathbf{a}_{ij}, \quad \mathbf{a} = \sum_\tau a + o_\tau \otimes \mathbf{s}_\tau \tag{1.17}$$
$$\mathbf{D} = \mathbf{d} \odot \mathbf{d}_0^{\odot -1}, \quad \mathbf{d}_{0i} = \sum_i \mathbf{d}_i, \quad \mathbf{d} = d + \mathbf{s}_1$$

In these sorts of simulation, the agent succumbs to the epistemic affordance of the instructional cues, until it learns that the reward is always on the left-hand side. At which point, the expected utility (or extrinsic value) of going directly to the baited arm exceeds the epistemic affordance (or intrinsic value) of soliciting the instructional cue. At this point, there is a switch from explorative to exploitative behaviour – the behavioural measure we used to compare sophisticated and unsophisticated schemes.

---

[13] Note that in order to accumulate beliefs about the context from trial to trial, it is necessary to carry over posterior beliefs about context from one trial as prior beliefs for the next (in the form of Dirichlet concentration parameters). For consistency with earlier formulations of this paradigm, we carry over the beliefs about the initial state on the previous trial that are evaluated using a conventional backwards pass; namely, the normalised likelihood of any given initial state, given subsequent observations – and probability transitions based on realised action.

## Exploration and exploitation in a T-maze

Figure 3 shows the results of three simulations. In these simulations, the rat performed 32 trials where each trial comprised two moves, starting from the central location. The prior preferences for reward and punishment outcomes were specified with the prior costs (**C**) of -2 and 2, respectively[14]. In these and subsequent simulations, actions were selected as the most likely (maximum *a posteriori*) action. Therefore, all subsequent simulations are deterministic realisations of (Bayes) optimal behaviour based upon expected free energy. The simulations in Figure 3 start with a sophisticated agent with a planning horizon of two (this corresponds to the depth of action sequences considered into the future). In other words, it accumulates the expected free energy for all plausible paths, until the end of each trial. This enables a confident and definitive epistemic policy selection that gives way to exploitation, when the rat realises the reward is always located in the left arm.

If we compare this performance with that of an unsophisticated rat – who just looks one move ahead – we see a similar behaviour. However, there are two differences. First, the rat is less confident about its behaviour, because it does not evaluate the consequences of its actions in terms of belief updating. Although it finds the instructional cue more attractive – in virtue of its epistemic affordance – it is still partially compelled to remain at the central location, which ensures that it will avoid aversive outcomes. Because the unsophisticated agent underestimates the epistemic affordance of the instructional cue, it paradoxically performs better in terms of suspending its information foraging earlier – and switching to exploitative behaviour a few trials before the sophisticated agent (but see below).

For completeness, we show the results of an unsophisticated agent, whose behaviour is predicated on Bayesian risk, i.e., with no epistemic value in play. As might be anticipated, this agent exposes itself to Bayesian risk – by foregoing a visit to the right or left arm – in a way that is precluded by agents who minimise expected free energy. Here, the starting and instructional cue locations are equally attractive. When the rat is lucky enough to select the lower arm, it knows what to do; however, it has no sense that this is the right kind of behaviour. After a sufficient number of trials, it realises that the reward is always on the left-hand side and starts to respond in an exploitative fashion, albeit with relatively low confidence. These results highlight the distinction between sophisticated and unsophisticated agents who predicate their policy selection on expected free energy – and between unsophisticated agents using expected free energy with and without epistemic affordance.

---

[14] Because costs are specified in terms of self-information or surprisal, they have meaningful and quantitative units. For example, a differential cost of three natural units corresponds to a log odds ratio 1:20 and reflects a strong preference for one state what outcome over another. This is the same interpretation of Bayes factors in statistics: Kass, R.E., Raftery, A.E., 1995. Bayes Factors. Journal of the American Statistical Association 90, 773-795. Here, the difference between reward and punishment was four natural units.

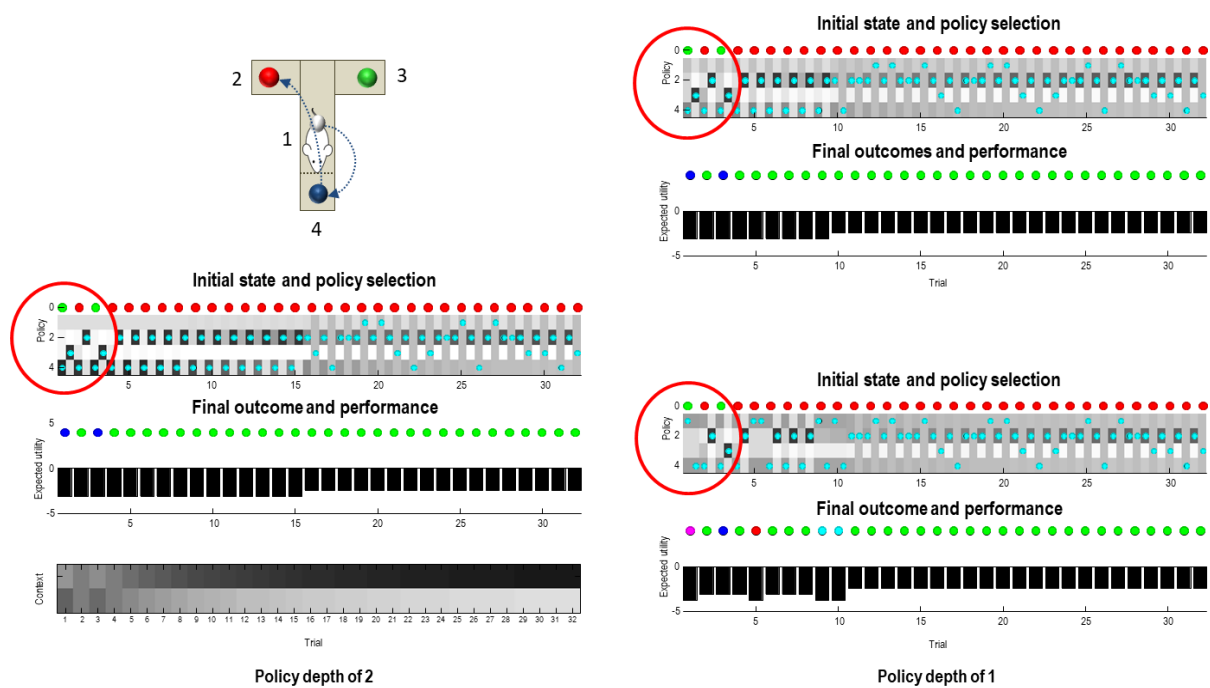

**FIGURE 3**

**Epistemic foraging in a T-maze.** This figure shows the results of simulations based upon the T-maze paradigm described in the main text. The left panel shows the results of simulating 32 trials, where the rat started at the central location. Each trial comprises two moves. The insert on the upper left illustrates foraging for information by interrogating the instructional cue in the lower arm and then securing the reward in the left arm. The results in each of the three panels have the same format. The upper row illustrates the predictive distribution over actions (moves to the central location, the left, the right and lower arm, respectively). The darker the colour, the more likely the action. The cyan dots are the actions that were sampled and executed at each epoch, within each trial. The coloured dots above indicate the hidden context; namely, whether the left or right arm was baited. The middle panel shows the resulting performance, in terms of the expected utility or negative Bayesian risk. The coloured circles show the final outcome (blue location 3 – right arm – and green location 2 – left arm). The lowest panel (on the left) shows the posterior beliefs about the hidden context (right versus left) based upon Dirichlet concentration parameters, accumulated over trials. The left panel of results shows confident epistemic behaviour with a planning horizon of two. As is typical in these kinds of simulations, the agent starts off by foraging for information and responding to the epistemic affordance of the instructional cue in the lower arm. However, because the reward is always encountered in the left arm (after the first couple of trials), the rat loses interest in the instructional cue – as it becomes more confident about where the reward is located. This experience-dependent loss of epistemic affordance leads to a switch from exploratory to exploitative behaviour; here at trial 16. A similar kind of behaviour is shown in the upper right panels; however, here, the planning horizon was reduced to one. In other words, the rat only considered the expected free energy of one move ahead. The key difference here is a less confident (i.e., precise) belief distribution over early actions (highlighted by the red circles). Although the lower arm has the greatest posterior probability, there is a nontrivial probability that the rat thinks it should stay where it is. This mild ambiguity about what should be done, means that exploratory behaviour yields to exploitative behaviour slightly earlier, at trial 10. Finally, the lower right panels show the results when expected free energy is replaced by Bayesian risk. In other words, any epistemic affordance of the instructional cue is precluded. This renders the posterior probability of staying or moving to the lower arm the same. When, by chance, the instructional cue is encountered, exploitative behaviour follows; however, there are times when the rat simply stays at the central location and learns nothing about the prevailing context. Note, that in this example, there are costly trials, in which the rat fails to visit either baited arm.

In the above simulations, the sophisticated agent persevered with its epistemic behaviour for longer than the unsophisticated agent. At first glance, this may seem to be a paradoxical result – if we were measuring performance in terms of Bayesian risk. However, this is not the case as illustrated in Figure 4. Here, we repeated the simulations above but with one small change: we made the epistemic cue mildly aversive, by giving it a cost of one. This has no effect on the sophisticated agent; other than slightly abbreviating the exploratory phase of activity. However, the unsophisticated agent has, understandably, been caught in a bind. The starting location is now marginally more preferable than the instructional cue – and it has no reason to leave the centre of the maze. While this ensures aversive outcomes are avoided, it also precludes epistemic foraging and subsequent exploitation. Heuristically, only the sophisticated agent can see past the short-term pain for the long-term gain. We will pursue this theme in the final simulations, where the agent's planning horizon becomes non-trivial.

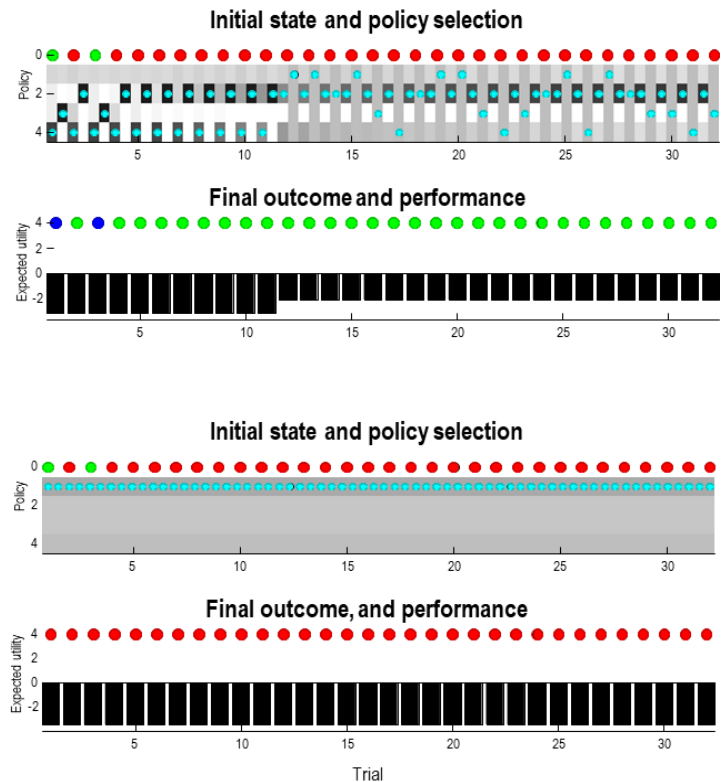

This reproduces the results of the previous figure with a deep policy search (of horizon or depth 2). However, here, we have made the lower arm slightly aversive. This is no problem for the sophisticated agent who 'sees through' the short-term cost to visit the instructional cue as usual. Because this location is mildly aversive, the switch to exploitative behaviour is now slightly earlier (at trial 12). Contrast this behaviour with an unsophisticated agent that does not look beyond its next move. The resulting behaviour is shown in the lower panels. Unsurprisingly, the agent just stays at the starting position and learns nothing about its environment – and safely avoids all adverse outcomes at the expense of forgoing any rewards.

## Deep planning and navigation

The simulations above show that a sophisticated belief-updating scheme enables more confident and nuanced policy selection – that translates into more efficient exploitative behaviour. To illustrate how this scheme scales up to deeper policy searches, we revisit a problem that has been previously addressed using a bespoke prior, based upon the graph Laplacian (Kaplan and Friston, 2018). This problem was previously framed in terms of navigation to a target location in a maze. Here, we forego any special priors to see if the sophisticated scheme could handle deep tree searches that underwrite paradoxical behaviours, like moving away from a target to secure it later (c.f., the mountain car problem). Crucially, in this instance, there was no ambiguity about the hidden states. However, there was ambiguity or uncertainty about the likelihood mapping that determines whether a particular location should be occupied or not. In other words, this example uses a more conventional foraging setup; in which the rat has to learn about the structure of the maze while, at the same time, pursuing its prior preferences to reach a target location. Here, exploratory behaviour is driven by the intrinsic value or information gain afforded to beliefs about parameters of the likelihood model – as opposed to hidden states. Colloquially, one can think of this as epistemic affordance that is underwritten by *novelty* as opposed to *salience* (Barto et al., 2013; Parr and Friston, 2019a; Schwartenbeck et al., 2019). Having said this, we anticipated that exactly the same kind of behaviour would arise and that the sophisticated scheme would be able to plan to learn – and then exploit what it has learned.

In this paradigm, a rat has to navigate over the 8×8 grid maze, where each location may or may not deliver a mildly aversive stimulus (e.g., a foot-shock). Navigation is motivated by prior preferences to occupy a target location; here, the centre. In the simulations below, the rat starts at the entrance to the maze and has a prior preference for safe outcomes (cost of -1) and against aversive outcomes (cost of +1). Prior preferences for location depend on the distance from the current position to the target location. The generative model for this setup is simple: there was one hidden factor with 64 states corresponding to all possible locations. These hidden states generate safe or aversive (somatosensory) outcomes, depending upon the location. In addition, (exteroceptive) cues are generated that directly report grid location. The five allowable actions comprised one step in any direction or staying put.

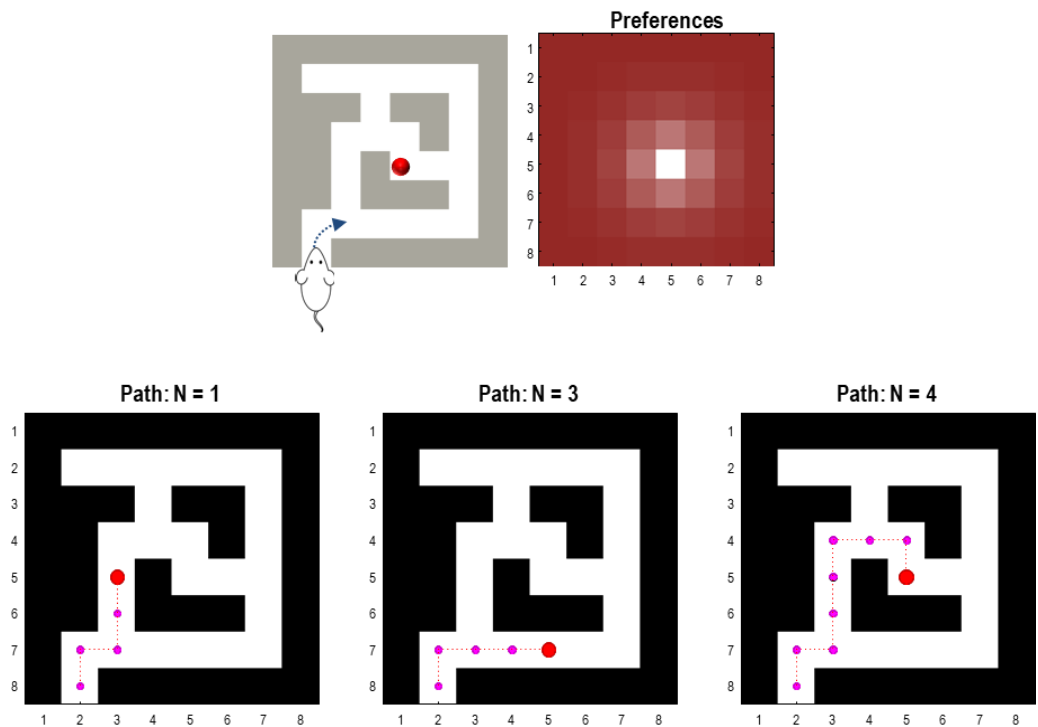

Figure 5

**Navigation as inference**. This figure reports the result of a simulated maze navigation. The upper panels illustrate the form of this maze, which comprises an 8 x 8 grid. Each location may or may not deliver a mildly aversive outcome (e.g., foot-shock). At the same time, the rat's prior preference is to be near the centre of the maze. These prior preferences are shown in image format in the upper right panel, where the log prior preference is illustrated in pink, with white being the most preferred location. The lower three panels record the trajectory or path taken by a rat from the starting location on the lower left. The three panels show the (deterministic) solutions for a planning horizon of one, three and four). With horizons of less than four, the rat gets stuck on the other side of an aversive barrier that is closest to the central (i.e., target) location. This is because any move away from this location – with a small excursion – has a smaller expected free energy than staying put. However, if the policy search is sufficiently deep (i.e., a planning horizon greater than three) then the rat can effectively imagine what would happen if it pressed deeper into the future; enabling long-term gains to supervene over short-term losses. This results in the rat inferring and pursuing the shortest path to the target location; even though it occasionally moves away from the centre. In these simulations, the rat knew the locations of the aversive outcomes and was motivated by minimising Bayesian risk.

Figure 5 shows the results of typical simulations, when increasing the planning horizon from one through to four. The key thing to observe here is that there is a critical horizon, which enables our subject to elude local minima of expected free energy as it pursues its goal. In these simulations, our subject was equipped with full knowledge of the aversive locations and simply planned a route to its target location. However, relatively unsophisticated agents get stuck on the other side of aversive barriers that are closest to the target location. In other words, they remain in locations in which the expected free energy of leaving is always greater than staying put (Cohen et al., 2007). This can happen when the planning horizon is insufficient to enable the rat to contemplate distal (and potentially preferable) outcomes (as seen in the lower left and middle panels of Figure 5). However, with a planning horizon of four (or more), these local minima are vitiated, and the rat easily plans – and executes – the shortest path to the target. In these simulations, the total number of moves was eight, which is sufficient to reach the target via the shortest path. This sort of behaviour is reminiscent of the prospective planning required to solve things like the mountain car problem. In other words, the path of least expected free energy can often involve excursions through state (and belief) space that point away from the ultimate goal.

Above, we characterised the degree of sophistication in terms of planning as inference. In this setting, there was no ambiguity about outcomes that would license an explanation in terms of epistemic affordance or salience – of the sort that motivated behaviour in the T-maze examples of the preceding section. However, we can reintroduce epistemics by introducing uncertainty about the locations that deliver aversive outcomes. Exploration now becomes driven by curiosity about the parameters of the likelihood mapping: see (1.9). One can illustrate the minimisation of expected free energy in terms of curiosity and novelty (Barto et al., 2013; Schmidhuber, 2006) by simulating a rat who has never been exposed to the maze previously. This was implemented by setting the prior (Dirichlet) parameters of the likelihood mapping between hidden states and somatosensory outcomes to a small value (i.e., 1/64). In terms of sufficient statistics, the expected free energy is now supplemented with a novelty term based upon posterior expectations about the likelihood mapping (Friston et al., 2017b):

$$\mathbf{G}(u_{\tau-1}, o_{\tau-1}) = \underbrace{\mathbf{s}_\tau^u \cdot [\ln \mathbf{s}_\tau^u + \mathbf{C} + \mathbf{H}]}_{\text{Next action}} - \underbrace{\mathbf{o}_\tau^u \cdot \mathbf{W} \mathbf{s}_\tau^u}_{\text{Novelty}} + \underbrace{\mathbf{u}_\tau^o \cdot \mathbf{G}(u_\tau, o_\tau) \mathbf{o}_\tau^u}_{\text{Subsequent actions}} \qquad (1.18)$$

$$\mathbf{W} = \tfrac{1}{2}(\mathbf{a}^{\odot-1} - \mathbf{a}_0^{\odot-1})$$

In addition, we removed preferences for a particular location in order to study purely exploratory behaviour. The results of the ensuing simulation are shown in Figure 6. In this example, the rat was allowed to make 64 consecutive moves, while updating the Dirichlet parameters after every move. The upper panels of Figure 6 show the resulting trajectory. The key thing to observe here is that nearly every location has been explored. This rests upon a trajectory in which previously visited locations lose their novelty or epistemic affordance;

thereby promoting policies that take the rat into uncharted territory. This kind of exploratory behaviour disappears if we replace expected free energy with Bayesian risk. In this setting, after the first move, the rat returns to its original location and just sits there for 64 trials (see the lower panels of Figure 6).

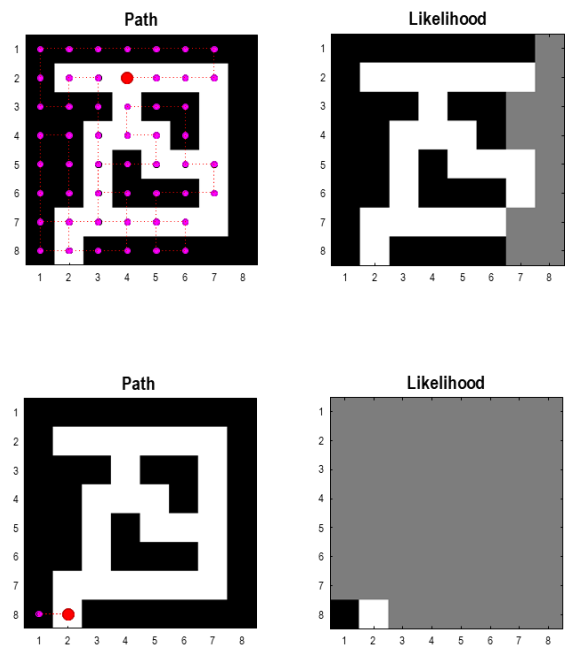

<div align="center">

FIGURE 6

</div>

**Exploration and novelty**. This figure reports the results of a simulation in the same maze as in the previous figure. However, here, we removed prior knowledge about which locations should be avoided – and prior preferences for being near the centre. This means that the only incentives for movement are purely epistemic in nature; namely, curiosity or the novelty of finding out "what would happen if I did that". This produces a trajectory of moves that explore the locations, building up a picture of where aversive (foot-shock) stimuli are elicited and where they are not. The key aspect of this trajectory is that it avoids revisiting previously explored locations – to provide a nearly optimal coverage of the exploration space. The number of moves was 64 (with an updating of the posterior beliefs about likelihood parameters after each move). This means that, in principle, the rat could have visited every location. Indeed, nearly every location has been visited – as shown on the upper right, in terms of the final likelihood of receiving an aversive stimulus at each location. The lower panels show the same results but after replacing expected free energy (that includes the novelty term) with Bayesian risk (that does not). Unsurprisingly, the Bayesian risk agent has no imperative to move, because it has no preferences about its location and, after the first move, realises it is in a safe location. In other words, after the first move, it returns to the starting location and remains there for the remainder of available trials. As such, it learns nothing about the mapping between location and sensory outcomes.

Finally, to simulate curiosity under a task set, we reinstated prior preferences about location. In this simulation, the rat has to resolve the dual imperative to satisfy its curiosity, while – at the same time – realise preferences for being at the centre of the maze. In other words, it has to contextualise its goal seeking behaviour in relation to what it knows about how to realise those goals. Figure 7 shows the results of a simulation in which the rat was given five exposures to the maze, each comprising eight moves with a planning horizon of four. Within four exposures, it has learned what it needs to learn – about the aversive locations – to plan the shortest path to its target location and execute that path successfully (dotted black line in the left panel of Figure 7). In contrast to Figure 6, the exploration is now limited to preferred locations with precise likelihood mappings that are sufficient to encompass the shortest path (compare the left panels of Figure 6 and Figure 7).

This completes our numerical analyses, in which we have looked at deep policies searches predicated on expected free energy, where expected free energy supplements Bayesian risk with epistemic affordance – either in terms of salience (resolving uncertainty about hidden states) or novelty (resolving uncertainty about hidden model parameters).

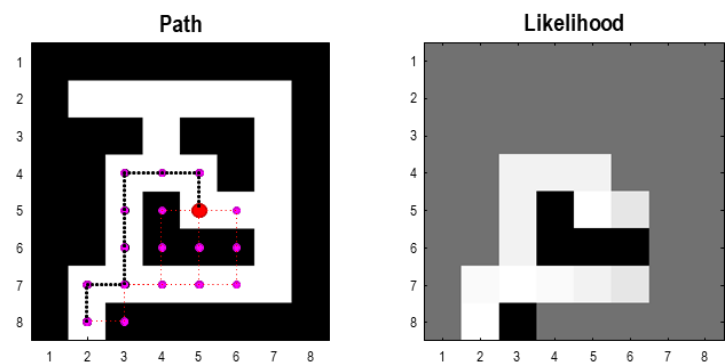

FIGURE 7

**Exploration under a task set**. This figure reproduces the same paradigm as in the previous figure but reinstating prior preferences about being near the centre of the maze (i.e., a task set). In this instance, the imperatives for action include both curiosity and pragmatic drives to realise prior preferences. The left panel shows a sequence of trajectories over five trials, where the rat was replaced at the initial location following eight moves. The right panel shows the final accumulated Dirichlet counts depicting the probability of an aversive outcomes at each location. This accumulated evidence – or familiarity with the environment – enables the rat to plan the shortest path to its target after just four exposures. This path is shown as the black dashed line in the left panel. Compare the likelihood mapping with the previous figure. Here, the agent restricted its exploration to those parts of the maze that encompass the path to its goal.

## Conclusion

This paper has described a recursive formulation of expected free energy that, effectively, instigates a deep tree search for planning as inference. The ensuing planning is sophisticated, in the sense that it entails beliefs about beliefs – in virtue of accumulating predictive posterior expectations of expected free energies down plausible paths. In other words, instead of just propagating beliefs about the consequences of successive actions, the scheme simulates belief updating in the future, based upon preceding beliefs about the consequences of action. This scheme was illustrated using a simple T-maze problem and a navigation problem that required a deeper search.

In the introduction, we noted that active inference may be difficult to scale, although remarkable progress has been made in this direction recently using amortized inference and sampling. For example, (Ueltzhöffer, 2018) parameterized both the generative model and approximate posterior with function approximators – using evolutionary schemes to minimise variational free energy, when gradients were not available. Similarly, (Millidge, 2019) amortized perception and action by learning a parametrised approximation to expected free energy. (Catal et al., 2019) focused on learning prior preferences, using a learning-from-example approach. (Tschantz et al., 2019) extended previous point-estimate models to include full distributions over parameters. This allowed them to apply active inference to continuous control problems (e.g., the mountain car problem, the inverted pendulum task and a challenging hopper task) and demonstrate an order of magnitude increase in sampling efficiency, relative to a strong model-free baseline (Lillicrap et al., 2015). Please see (Tschantz et al., 2019) for a full discussion – and a useful deconstruction of active inference, in relation to things like model-based reinforcement learning; e.g., (Schrittwieser et al., 2019).

Note that the navigation example is an instance of planning to learn. As such, it solves the kinds of problems for which reinforcement learning and its variants usually address. In other words, we were able to solve a learning problem from first (i.e. variational) principles, without recourse to backwards induction or other (belief-free) schemes like Q-learning, SARSA or successor representations; e.g., (Dayan, 1993; Gershman, 2017; Momennejad et al., 2017; Russek et al., 2017). This is potentially important because predicating an optimisation scheme on inference – as opposed to learning – endows it with a context sensitivity that eludes many learning algorithms (Daw et al., 2011). In other words, because there are probabilistic representations of time-sensitive hidden states (and implicit uncertainty about those states), behaviour is motivated by resolving uncertainty about the context in which an agent is operating. This may be the kind of (Bayesian) mechanics that licenses the notion of competent schemes that can both 'learn to plan' and 'plan to learn'.

The current formulation of active inference does not call on sampling or matrix inversions – the Bayes optimal belief-updating deals with uncertainty in a deterministic fashion. Conceptually, this reflects the difference between the stochastic aspects of random dynamical systems and the deterministic behaviour of the accompanying density dynamics, which describe the probabilistic evolution of those systems (e.g., the Fokker Planck equation). Because active inference works in belief spaces, i.e., on statistical manifolds (Da Costa et al., 2020b). There is no need for sampling or random searches – the optimal paths are, instead, evaluated by propagating beliefs or probability distributions into the future to find the path of least variational free energy (Friston, 2013).

In the setting of deep policies searches, this approach has the practical advantage of terminating searches over particular paths, when they become implausible. For example, in the navigation example above, there were five actions and 64 hidden states leading to a large number of potential paths ($1.0486 \cdot 10^{10}$ for a planning horizon of four and $1.0737 \cdot 10^{15}$ for a planning horizon of six). However, only a tiny fraction of these paths is actually evaluated – usually several hundred, which takes a few hundred milliseconds on a personal computer. Given reasonably precise beliefs about current states and state transitions, only a small number of paths are eligible for evaluation, which leads us to our final comment on the scalability of active inference.

## Limitations

In one sense, we have addressed scaling through the computational efficiency afforded by belief propagation using the sophisticated scheme above. However, we have only illustrated this scheme on rather trivial problems. In principle, one can scale up the dimensionality of state spaces (and outcomes) with a degree of impunity. This follows from the fact that the number of plausible states (and transitions) can be substantially constrained, using the right kind of generative model – that leverages factorisations and sparsity. For example, the factorisation between hidden states and actions used above rests upon the implicit assumption that every action is allowed from every state. This is a strong assumption but perfectly apt for many generative models.

One could also call upon a related symmetry; namely, a hierarchical separation of temporal scales in deep models, where one Markov decision process is placed on top of another (Friston et al., 2017d; George and Hawkins, 2009; Hesp et al., 2019b; Rikhye et al., 2019). In these models, transitions at the higher level usually unfold at a slower time scale than the level below. This engenders semi-Markovian dependencies that can generate complicated and structured behaviours. In this setting, one could consider hidden states at higher levels that generate the initial and final states of the level below. Policy optimisation within each level – using a sophisticated scheme – could then realise the trajectory between the initial states (i.e., empirical priors over initial states) and final states (i.e., priors that determine the cost function and subsequent empirical priors over action).

Finally, it should be noted that in many applications, the states and actions of real world processes are continuous, which presents a further scaling challenge for discrete state space models (Tschantz et al., 2019). However, it is possible to combine sophisticated (discrete) schemes with continuous models; provided one uses the appropriate message passing between the continuous and discrete levels. For example, (Friston et al., 2017c) used a Markov decision process to drive continuous eye movements. Indeed, it would be interesting to revisit simulations of saccadic searches using sophisticated inference, especially in the context of reading.

# Appendix

This appendix considers two lemmas that underwrite expected free energy from two complementary perspectives. The first is based on a generative model that combines the principles of optimal Bayesian design (Lindley, 1956) and decision theory (Berger, 2011), while the second is based upon a principled account of self-organisation (Friston, 2019; Parr et al., 2020). Finally, we consider several corollaries that speak to the notions of active inference (Friston et al., 2015), empowerment (Klyubin et al., 2005), information bottlenecks (Tishby et al., 1999), self-organisation (Friston, 2013) and self-evidencing (Hohwy, 2016). In what follows, $Q(o_\tau, s_\tau, \pi)$ denotes a predictive distribution over future variables and policies, conditioned upon initial observations; while $P(o_\tau, s_\tau, \pi)$ denotes a generative model; i.e., a marginal distribution over final states and policies. For simplicity, we will omit model parameters and assume policies start from the current time point, allowing us to omit the variational free energy from the generalised free energy (since observational evidence is the same for all policies).

**Objective**: our objective is to establish a generalised free energy functional that can be minimised with respect to a posterior over policies; noting that this posterior is necessary to marginalise the joint posterior over hidden states and policies to infer hidden states. To comply with Bayesian decision theory, generalised free energy can be constructed to place an upper bound on *Bayesian risk*, which corresponds to the divergence between the predictive distribution over outcomes and prior preferences. In other words, Bayesian risk is the expected *surprisal* or negative log evidence. To comply with optimal Bayesian design, one can specify priors over policies that lead to states with a precise likelihood mapping to observable outcomes.

**Lemma**: (*Bayes optimality*): generalised free energy[15] is an upper bound on risk, under a generative model whose priors over policies lead to states with precise likelihoods:

$$F[Q(s, \pi)] = \underbrace{\mathbb{E}_Q[\ln Q(s_\tau, \pi) - \ln P(o_\tau, s_\tau)]}_{\text{Generalised free energy}} \geq \underbrace{D_{KL}[Q(o_\tau) \| P(o_\tau)]}_{\text{Risk}}$$

$$\log P(\pi) = \underbrace{\mathbb{E}_{Q(o_\tau, s_\tau|\pi)}[\log P(o_\tau \mid s_\tau)]}_{\text{Empirical prior}}$$

$$(1.19)$$

Note that $P(\pi)$ is an *empirical prior* because it depends upon the predictive density that depends upon past observations. The priors over hidden states and outcomes can be regarded as a target distribution or *prior preferences*.

**Proof**. By substituting the empirical prior (1.19) into the expression for free energy, we have (noting that policies and outcomes are conditionally independent, given hidden states):

---

[15] Equation (1.19) follows from (1.1), when treating $F(\pi)$ and $E(\pi)$ as constants; i.e., ignoring past observations and empirical priors over policies.

$$F[Q(s,\pi)] = \underbrace{\mathbb{E}_Q[D_{KL}[Q(s_\tau \mid \pi) \| P(s_\tau)]]}_{\text{Expected risk (States)}} + \underbrace{D_{KL}[Q(\pi) \| P(\pi)]}_{\text{Complexity (Policies)}}$$

$$\geq \underbrace{\mathbb{E}_Q[D_{KL}[Q(s_\tau \mid \pi) \| P(s_\tau)]]}_{\text{Expected risk (States)}}$$

$$= \underbrace{\mathbb{E}_Q[D_{KL}[Q(o_\tau \mid \pi) \| P(o_\tau)]]}_{\text{Expected risk (Outcomes)}} + \underbrace{\mathbb{E}_Q[D_{KL}[Q(s_\tau \mid o_\tau, \pi) \| P(s_\tau \mid o_\tau)]]}_{\text{Expected evidence bound}}$$

$$\geq \underbrace{\mathbb{E}_Q[D_{KL}[Q(o_\tau \mid \pi) \| P(o_\tau)]]}_{\text{Expected risk (Outcomes)}}$$

$$= \underbrace{D_{KL}[Q(o_\tau) \| P(o_\tau)]}_{\text{Risk}} + \underbrace{\mathbb{E}_Q[D_{KL}[Q(o_\tau \mid \pi) \| Q(o_\tau)]]}_{\text{Mutual information}}$$

$$\geq \underbrace{D_{KL}[Q(o_\tau) \| P(o_\tau)]}_{\text{Risk}}$$

(1.20)

These inequalities show that generalised free energy upper bounds the predictive divergence from the marginal likelihood over outcomes, i.e., model evidence. When this bound is minimised; (i) the complexity cost of policies is minimised – enforcing prior beliefs about policies; (ii) the predictive posterior over hidden states becomes the posterior under the generative model and (iii) policies and outcomes become independent. This independence follows by construction of the free energy functional and means that final outcomes do not depend upon initial conditions; implying a form of steady-state (please see below) □

**Corollary** (*expected free energy*): the free energy can now be minimised w.r.t. the posterior over policies, by expressing free energy in terms of expected free energy:

$$F[Q(s,\pi)] = \mathbb{E}_{Q(\pi)}[\mathrm{G}(\pi) + \ln Q(\pi)]$$

$$Q(\pi) = \arg\min_Q F[Q(s,\pi)] \Rightarrow -\ln Q(\pi) = \mathrm{G}(\pi)$$

(1.21)

$$\mathrm{G}(\pi) = \mathbb{E}_{Q(o_\tau, s_\tau \mid \pi)} \underbrace{[\ln Q(s_\tau \mid \pi) - \ln P(o_\tau, s_\tau)]}_{\text{Expected free energy}}$$

$$= \underbrace{D_{KL}[Q(s_\tau \mid \pi) \| P(s_\tau)]}_{\text{Expected risk}} - \underbrace{\mathbb{E}_{Q(o_\tau, s_\tau \mid \pi)}[\ln P(o_\tau \mid s_\tau)]}_{\text{Expected ambiguity}}$$

This renders free energy $F[Q(s,\pi)] = \mathbb{E}_Q[\mathrm{G}(\pi)] - \mathrm{H}[Q(\pi)]$ an expected energy minus the entropy of the posterior over policies, in the usual way. Finally, we can express the expected free energy of a policy as a bound on information gain and Bayesian risk:

$$G(\pi) = \underbrace{\mathbb{E}_Q[D_{KL}[Q(s_\tau \mid o_\tau, \pi) \parallel P(s_\tau \mid o_\tau)]]}_{\text{Expected evidence bound}} - \underbrace{\mathbb{E}_Q[\ln P(o_\tau)]}_{\text{Expected log evidence}}$$

$$- \underbrace{\mathbb{E}_Q[D_{KL}[Q(s_\tau \mid o_\tau, \pi) \parallel Q(s_\tau \mid \pi)]]}_{\text{Expected information gain}} \qquad (1.22)$$

$$\geq -\underbrace{\mathbb{E}_Q[D_{KL}[Q(s_\tau \mid o_\tau, \pi) \parallel Q(s_\tau \mid \pi)]]}_{\text{Expected information gain}} - \underbrace{\mathbb{E}_Q[\ln P(o_\tau)]}_{\text{Bayesian risk}}$$

This inequality shows that the free energy of a policy upper bounds a mixture of its expected information gain (Lindley, 1956) and Bayesian risk (Berger, 2011), where Bayesian risk is expected log evidence □

**Remarks**: Here, policies are treated as random variables, which means that planning as inference (Attias, 2003; Botvinick and Toussaint, 2012) becomes belief updating under optimal Bayesian design priors (Lindley, 1956; MacKay, 1992). One might ask what licenses these priors above? Although they can be motivated in terms of information gain (1.22), there is a more straightforward motivation that arises as a steady-state solution. We now turn to this complementary perspective that inherits from the Bayesian mechanics described in (Friston, 2019). Here, we are interested in situations when the predictive distribution attains its steady-state or target distribution.

**Objective**: We seek distributions over policies that afford steady-state solutions, i.e., when the final distribution does not depend upon initial observations. Such solutions ensure that, on average, stochastic policies lead to a steady-state or target distribution specified by the generative model. These solutions exist in virtue of conditional independencies, where the hidden states provide a Markov blanket (c.f., information bottleneck) that separates policies from outcomes. In other words, policies cause final states that cause outcomes. We will see below that there is a family of such solutions, where the Bayes optimality solution above is a special (canonical) case. In what follows, $Q(o_\tau, s_\tau, \pi) \coloneqq P(o_\tau, s_\tau, \pi \mid o_\leq)$ can be read as a posterior distribution, given initial conditions.

**Lemma** (*nonequilibrium steady-state*): When the surprisal of policies corresponds to a Gibbs free energy $G(\pi, \beta)$, the final distribution attains steady-state:

$$-\log Q(\pi) = \mathrm{G}(\pi, \beta) \Rightarrow D_{KL}[Q \parallel P] = 0$$

$$\mathrm{G}(\pi,\beta) = \underbrace{D_{KL}[Q(s_\tau \mid \pi) \parallel P(s_\tau)]}_{\text{Expected risk}} \underbrace{-\mathbb{E}_{Q(o_\tau, s_\tau \mid \pi)}[\beta \log P(o_\tau \mid s_\tau)]}_{\text{Expected ambiguity}}$$

$$\beta = \frac{\mathbb{E}_Q[\ln Q(\pi \mid s_\tau)]}{\mathbb{E}_Q[\ln P(o_\tau \mid s_\tau)]} = \frac{\mathrm{H}(\Pi \mid S_\tau, o_\leq)}{\mathrm{H}(O_\tau \mid S_\tau)}$$

$$P = Q(\pi \mid s_\tau) P(o_\tau, s_\tau)$$

$$Q = P(o_\tau \mid s_\tau) Q(s_\tau, \pi)$$

(1.23)

Here, $\beta \geq 0$ characterises steady-state in terms of the relative precision of policies and final outcomes, given final states. The generative model stipulates steady-state, in the sense that distribution over final states (and outcomes) does not depend upon initial observations. Here, the generative and predictive distributions simply express the conditional independence between policies and final outcomes, given final states. Note that when $\beta = 1$, Gibbs free energy becomes expected free energy.

**Proof**: rearranging the expressions in (1.23) and substituting into the KL divergence between the predictive and generative distributions gives:

$$\mathbb{E}_Q\left[\ln Q(\pi \mid s_\tau)\right] = \mathbb{E}_Q\left[\beta \ln P(o_\tau \mid s_\tau)\right] \Rightarrow$$

$$\mathbb{E}_Q\left[\mathrm{G}(\pi, \beta)\right] = \mathbb{E}_Q\left[\ln \frac{Q(s_\tau \mid \pi)}{Q(\pi \mid s_\tau) P(s_\tau)}\right] = -\mathbb{E}_Q[\ln Q(\pi)]$$

$$\Rightarrow$$

$$D_{KL}\left[Q(o_\tau, s_\tau, \pi) \parallel P(o_\tau, s_\tau, \pi)\right] = \mathbb{E}_Q\left[\ln \frac{Q(s_\tau \mid \pi)}{Q(\pi \mid s_\tau) P(s_\tau)} + \ln \frac{P(o_\tau \mid s_\tau)}{P(o_\tau \mid s_\tau)} + \ln Q(\pi)\right] = 0$$

(1.24)

This solution describes a particular kind of steady-state, where policies lead to (steady) states with more or less precise likelihoods, depending upon the value of $\beta$ □

**Remarks**: At steady-state, hidden states (and outcomes) 'forget' about initial observations, placing constraints on the distribution over policies that can be expressed in terms of a Gibbs free energy. In the limiting case that $\beta = 0$ we obtain a simple steady state where:

$$\mathrm{G}(\pi, 0) = \mathbb{E}_{Q(o_\tau, s_\tau \mid \pi)}\left[\ln \frac{Q(s_\tau \mid \pi)}{P(s_\tau)}\right] = D_{KL}[Q(s_\tau \mid \pi) \parallel P(s_\tau)]$$

(1.25)

This solution corresponds to a standard stochastic control, variously known as KL control or risk-sensitive control (van den Broek et al., 2010). In other words, one picks policies that minimise the divergence between the predictive and target distribution. More generally (i.e., $\beta > 0$), policies are more likely when they lead to states with a precise likelihood mapping. One perspective – on the distinction between simple and general steady-states – is in terms of conditional uncertainty about policies. For example, simple (i.e., $\beta = 0$) steady-states preclude uncertainty about which policy led to a final state. This would be appropriate for describing classical systems (that follow a unique path of least action), where it would be possible to infer which policy had been pursued, given the initial and final outcomes. Conversely, in general steady-state systems (e.g., mice and men), simply knowing that 'you are here' does not tell me 'how you got here', even if I knew where you were this morning. Put another way, there are lots of paths or policies open to systems that attain a general steady state.

The treatment in (Friston, 2019) effectively turns the steady-state lemma on its head by assuming the steady-state in (1.23) is stipulatively true – and then characterises the ensuing self-organisation in terms of Bayes optimal policies. In active inference, we are interested in a certain class of systems that self-organise to general steady-states; namely, those that move through a large number of probabilistic configurations from their initial state to their final (steady) state. In terms of information geometry, this means that the *information distance* between any initial and the final (steady) state is large. In the current setting, we could replace *information distance* (Crooks, 2007; Kim, 2018) by *information gain* (Lindley, 1956; MacKay, 1992; Still and Precup, 2012). That is, we are interested in systems that attain steady-state (i.e., target distributions) with policies associated with a large information gain[16]. Although not pursued here, general steady states with precise likelihood mappings have precise Fisher information matrices and information geometries that distinguish general forms of self-organisation from simple forms (Amari, 1998; Ay, 2015; Caticha, 2015; Ikeda et al., 2004; Kim, 2018). This perspective can be unpacked in terms of information theory with the following corollaries, which speak to active inference, empowerment, information bottlenecks, self-organisation, and self-evidencing.

---

[16]Note that a divergence such as information gain is not a measure of distance. The information distance (a.k.a. information length) can be regarded as the accumulated divergences along a path on a statistical manifold from the initial location to the final location.

**Corollary** (*active inference*): If a system attains a general steady-state, then by the Bayes optimality lemma, it will appear to behave in a Bayes optimal fashion – both in terms of optimal Bayesian design (i.e., exploration) and Bayesian decision theory (i.e., exploitation). Crucially, the loss function defining Bayesian risk is the negative log evidence for the generative model entailed by an agent. In short, systems (i.e., agents) that attain general steady-states will look as if they are responding to epistemic affordances (Parr and Friston, 2017) □

**Corollary** (*empowerment*): At its simplest, *empowerment* (Klyubin et al., 2005) underwrites exploration (i.e., intrinsic motivation) by exploring as many states in the future as possible – and thereby keeping options open. This exploratory imperative is evinced clearly if we generalise free energy to include $\beta$:

$$
\begin{aligned}
F[Q(s,\pi)] &= \mathbb{E}_Q\left[\ln\frac{Q(s_\tau,\pi)}{P(o_\tau\mid s_\tau)^\beta P(s_\tau)}\right] = \mathbb{E}_Q\left[\ln\frac{Q(s_\tau\mid\pi)Q(\pi)}{P(s_\tau)Q(\pi\mid s_\tau)}\right] \\
&= \underbrace{D_{KL}[Q(s_\tau\mid\pi)\parallel P(s_\tau)]}_{\text{Risk}} - \underbrace{I(\Pi;S_\tau\mid o_\leq)}_{\text{Empowerment}}
\end{aligned}
\tag{1.26}
$$

This expresses the free energy of the predictive distribution over final states and policies in terms of risk and empowerment. Minimising free energy with respect to policies therefore maximises empowerment; namely, the mutual information between policies and their final states, given initial observations. The epistemic aspect of empowerment can be seen by expressing it in terms of expected ambiguity:

$$
\underbrace{I(\Pi;S_\tau\mid o_\leq)}_{\text{Empowerment}} = \underbrace{H(\Pi\mid o_\leq)}_{\text{Entropy}} - \underbrace{\mathbb{E}_Q[\beta\ln P(o_\tau\mid s_\tau)]}_{\text{Expected ambiguity}}
\tag{1.27}
$$

On this reading, empowerment corresponds to minimising expected ambiguity, while maximising the entropy of policies. In other words, keeping (policy) options open by avoiding situations from which there is only one 'escape route'. Note that empowerment is a special case of active inference, when we can ignore risk (i.e., when all policies are equally risky) □

**Corollary** (*information bottleneck*): the information bottleneck method and related formulations (Bialek et al., 2001; Still et al., 2012; Tishby et al., 1999; Tishby and Polani, 2010) can be seen as generalisations of rate distortion theory. On this view, we can consider hidden states as an information bottleneck (c.f., Markov blanket) that plays the role of a compressed representation of past outcomes that best predict future outcomes. Here, we can regard the policies as mapping between initial and final observations, via hidden states. The information bottleneck method provides an objective function that can be minimised with respect to the distribution over policies. This (information bottleneck) objective function can be expressed in terms of the expected Gibbs energy as follows:

$$\mathbb{E}_{P(o_\leq|\pi)}\big[G(\pi,\beta)\big] = \mathbb{E}_{P(o_\tau,s_\tau,o_\leq|\pi)}\left[\ln\frac{P(s_\tau\mid o_\leq,\pi)}{P(s_\tau)} + \beta\ln\frac{P(o_\tau)}{P(o_\tau\mid s_\tau)} - \beta\ln P(o_\tau)\right]$$
$$= \underbrace{I(O_\leq;S_\tau\mid\pi) - \beta I(S_\tau;O_\tau)}_{\text{Information bottleneck}} \underbrace{-\mathbb{E}_P[\beta\ln P(o_\tau)]}_{\text{Bayesian risk}}$$

(1.28)

This means the average Gibbs energy of a policy, over initial observations, combines the information bottleneck objective function and Bayesian risk. Minimising the first term of the objective function (i.e., the mutual information between initial outcomes and hidden states) plays the role of compression, while maximising the second (i.e., the mutual information between hidden states final and outcomes) ensures the information gain that characterises general steady-states. Indeed, when relative precision $\beta = 1$, it is straightforward to show that the information bottleneck is an upper bound on expected information gain:

$$\underbrace{I(O_\leq;S_\tau\mid\pi) - I(S_\tau;O_\tau)}_{\text{Information bottleneck}} = \mathbb{E}_{P(o_\tau,s_\tau,o_\leq|\pi)}[\ln Q(s_\tau\mid\pi) - \ln P(s_\tau\mid o_\tau)]$$
$$= -\mathbb{E}_{P(o_\tau|\pi)}[\underbrace{D_{KL}[Q(s_\tau\mid o_\tau,\pi)\,\|\,Q(s_\tau\mid\pi)]}_{\text{Expected information gain}} + \underbrace{D_{KL}[Q(s_\tau\mid o_\tau,\pi)\,\|\,P(s_\tau\mid o_\tau)]}_{\text{Expected evidence bound}}]]$$
$$\geq -\mathbb{E}_{P(o_\tau|\pi)}[\underbrace{D_{KL}[Q(s_\tau\mid o_\tau,\pi)\,\|\,Q(s_\tau\mid\pi)]}_{\text{Expected information gain}}]] = -I(S_\tau;O_\tau\mid O_\leq,\pi)$$

(1.29)

Because the information bottleneck objective function is an average over initial observations it cannot be used directly for online (active) planning as inference; however, it can be used to learn fixed outcome-action policies (Hafez-Kolahi and Kasaei, 2019; Tishby and Zaslavsky, 2015). Note that the information bottleneck method is a special case of active inference, when we can ignore Bayesian risk (i.e., when all policies are equally risky) □

**Corollary**: (*self-organisation*): the average of expected free energy over policies can be decomposed into risk and conditional entropy:

$$
\begin{aligned}
\mathbb{E}_{Q(\pi)}[\mathrm{G}(\pi)] &= \underbrace{\mathbb{E}_Q[\ln Q(s_\tau \mid \pi) - \ln P(o_\tau, s_\tau)]}_{\text{Expected free energy}} \\
&= \underbrace{\mathbb{E}_Q[D_{KL}[Q(s_\tau \mid \pi) \| P(s_\tau)]}_{\text{Expected risk}} + \underbrace{\mathbb{E}_Q[-\ln Q(o_\tau \mid s_\tau)]}_{\text{Expected ambiguity}} \\
&= \underbrace{\mathbb{E}_Q[D_{KL}[Q(s_\tau \mid \pi) \| P(s_\tau)]}_{\text{Expected risk}} + \underbrace{\mathrm{H}(O_\tau \mid S_\tau, o_{\leq})}_{\text{Conditional entropy}} \geq 0
\end{aligned}
\tag{1.30}
$$

This decomposition means that if the expected free energy of policies is – on average – small, the predictive distribution over hidden states will converge to the prior or preferred distribution, while uncertainty about consequent outcomes will be small. In the limit, the predictive distribution over hidden states becomes the prior distribution, with no uncertainty about outcomes. This can be read as the limiting case of self-organisation to prior beliefs □

**Corollary**: (*self-evidencing*): the average of expected free energy over policies furnishes an upper bound on the (negative) expected log evidence of outcomes and the mutual information between these outcomes and their causes (i.e., hidden states):

$$
\begin{aligned}
\mathbb{E}_{Q(\pi)}[\mathrm{G}(\pi)] &= \underbrace{\mathbb{E}_Q[\ln Q(s_\tau \mid \pi) - \ln P(o_\tau, s_\tau)]}_{\text{Expected free energy}} \\
&= -\underbrace{\mathbb{E}_{Q(o_\tau, \pi)}[D_{KL}[Q(s_\tau \mid o_\tau, \pi) \| Q(s_\tau \mid \pi)]]}_{\text{Expected information gain}} - \underbrace{\mathbb{E}_{Q(o_\tau)}[\ln P(o_\tau)]}_{\text{Expected log evidence}} \\
&\quad + \underbrace{\mathbb{E}_{Q(o_\tau, \pi)}[D_{KL}[Q(s_\tau \mid o_\tau, \pi) \| P(s_\tau \mid o_\tau)]]}_{\text{Expected evidence bound}} \\
&\geq -\underbrace{I(S_\tau, O_\tau \mid \Pi, o_{\leq})}_{\text{Mutual information}} - \underbrace{\mathbb{E}_{Q(o_\tau)}[\ln P(o_\tau)]}_{\text{Expected log evidence}}
\end{aligned}
\tag{1.31}
$$

This decomposition means that if the expected free energy of policies is – on average – small, the expected log evidence and the mutual information between predicted states and the outcomes they generate will be large. In the limit, expected log evidence is maximised, with no uncertainty about outcomes, given hidden states. This can be read as the limiting case of self-evidencing with unambiguous outcomes □

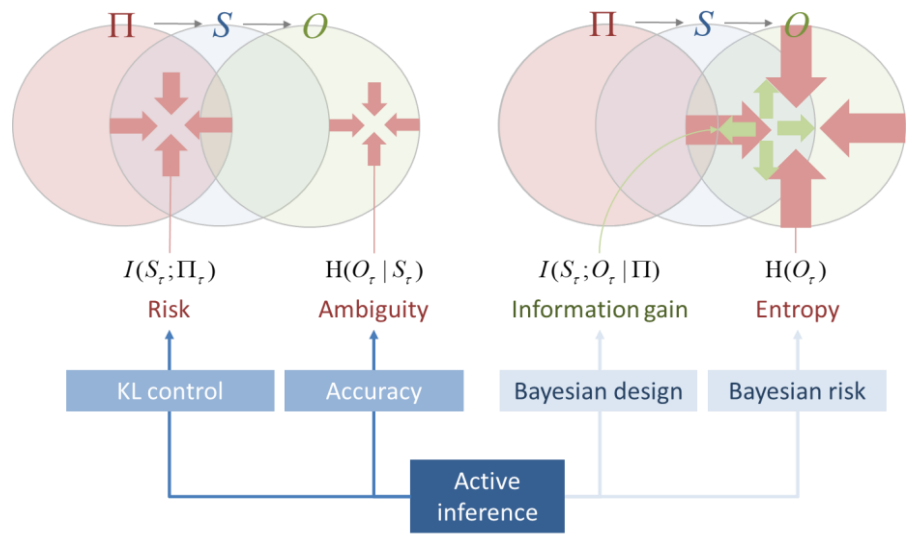

**Active inference and other schemes**. This schematic summarises the various imperatives implied by minimising a free energy functional of posterior beliefs about policies, ensuing states and subsequent outcomes. The information diagrams in the upper panels represent the entropy of the three variables, where intersections correspond to shared information or mutual information. A conditional entropy corresponds to an area that precludes the variable upon which the entropy is conditioned. Note that there is no overlap between policies and outcomes that is outside hidden states. This is because hidden states form a Markov blanket (i.e., information bottleneck) between policies and outcomes. Two complementary formulations of minimising expected free energy are shown on the right (in terms of risk and ambiguity) and left (in terms of information gain and entropy), respectively. One can see that both will tend to increase the overlap or mutual information between hidden states and outputs, while minimising entropy or Bayesian risk. In these diagrams, we have assumed steady-state, such that risk becomes the mutual information between policies and hidden states. For simplicity, we have omitted dependencies on initial observations. The various schemes or formulations considered in the text are shown at the bottom. These demonstrate that Bayesian decision theory (i.e., KL control and Bayesian risk) and optimal Bayesian design figure as complementary imperatives.

It can sometimes be difficult to see the relationships between the various conditional entropy and mutual information terms that constitute the free energy functional. Figure 8 tries to clarify these relationships using information diagrams. This schematic highlights the complementary decompositions of expected free energy in terms of risk and ambiguity – and information gain and entropy. These decompositions are summarised in terms of the imperative to minimise various segments of the information diagrams. Figure 1 then highlights the particular components that figure in special cases, such as an optimal Bayesian decisions and design.

## Software note

Although the generative model changes from application to application, the belief updates described in this paper are generic and can be implemented using standard routines (here **spm_MDP_VB_XX.m**). These routines are available as Matlab code in the SPM academic software: [http://www.fil.ion.ucl.ac.uk/spm/](http://www.fil.ion.ucl.ac.uk/spm/). The simulations in this paper can be reproduced (and customised) via a graphical user interface by typing >> **DEM** and selecting the appropriate (**T-maze or Navigation**) demo.

## Acknowledgements

KJF was funded by the Wellcome Trust (Ref: 088130/Z/09/Z). LD is supported by the Fonds National de la Recherche, Luxembourg (Project code: 13568875). CH was funded by a Research Talent Grant (no. 406.18.535) of the Netherlands Organisation for Scientific Research (NWO).

## Disclosure statement

The authors have no disclosures or conflict of interest.